\documentclass[]{aa}
\usepackage[T1]{fontenc}
\usepackage[utf8]{inputenx}
\usepackage{lmodern}
\usepackage{graphicx}
\usepackage{natbib}
\bibpunct{(}{)}{;}{a}{}{,} 
\setcitestyle{round}
\usepackage{ulem}
\usepackage{textcomp}
\usepackage{gensymb}
\usepackage{longtable}
\usepackage{threeparttable}
\usepackage{multicol}
\usepackage{multirow}
\usepackage{lipsum}
\usepackage{float}
\usepackage{todonotes}
\usepackage[Symbol]{upgreek}
\usepackage{amsmath}
\usepackage{etoolbox}
\usepackage{txfonts}
\usepackage{url}
\usepackage{xcolor}
\usepackage{mathtools}

\DeclarePairedDelimiter\abs{\lvert}{\rvert}

\usepackage[most]{tcolorbox}
\usepackage[colorlinks=true,allcolors=blue]{hyperref}
\usepackage{xcolor}
\newcommand{\enzo}{{\it {\small ENZO}}}

\begin{document}
 
\title{ Exploring the relation between turbulent velocity and density fluctuations in the stratified intracluster medium
}

\author{M. Simonte \inst{1,2}, F. Vazza \inst{1,2,3}, F. Brighenti \inst{1}, M. Br\"uggen \inst{2}, T. W. Jones \inst{4}, M. Angelinelli \inst{1,5}}

\institute{Dipartimento di Fisica e Astronomia, Universit\'{a} di Bologna, Via Gobetti 93/2, 40122, Bologna, Italy
\and  Hamburger Sternwarte, University of Hamburg, Gojenbergsweg 112, 21029 Hamburg, Germany
\and Istituto di Radioastronomia, INAF, Via Gobetti 101, 40122, Bologna, Italy
\and  School of Physics and Astronomy, University of Minnesota Twin Cities Minneapolis, MN, USA
\and INAF, Osservatorio di Astrofisica e Scienza dello Spazio, via Pietro Gobetti 93/3, 40129 Bologna, Italy}

 \authorrunning{M. Simonte et al.}
 \titlerunning{Exploring the relation between turbulence, gas fluctuations and gravity in the simulated intracluster medium}

\date{Accepted ???. Received ???; in original form ???}

\abstract
   {The dynamics of the intracluster medium (ICM) is affected by turbulence driven by several processes, such as mergers, accretion and feedback from active galactic nuclei.} 
   {X-ray surface brightness fluctuations have been used to constrain turbulence in galaxy clusters. Here, we use simulations to further investigate the relation between gas density and  turbulent velocity fluctuations, with a focus on the effect of the stratification of the ICM.}
   {In this work, we studied the turbulence driven by hierarchical accretion by analysing a sample of galaxy clusters simulated with the cosmological code ENZO. We used a fixed scale filtering approach to disentangle laminar from turbulent flows.}
   {In dynamically perturbed galaxy clusters, we found a relation between the root mean square of density and velocity fluctuations, albeit with a different slope than previously reported. 
    The Richardson number is a parameter that represents the ratio between turbulence and buoyancy, and we found that this variable has a strong dependence on the filtering scale.  However, we could not detect any strong relation between the Richardson number and the logarithmic density fluctuations,  in contrast to results by recent and more idealised simulations. In particular, we find a strong effect from radial accretion, which appears to be the main driver for the gas fluctuations. The ubiquitous radial bias in the dynamics of the ICM suggests that homogeneity and isotropy are not always valid assumptions, even if the turbulent spectra follow Kolmogorov's scaling.
    Finally, we find that the slope of the velocity and density spectra are independent of cluster-centric radii.}
    {}



\maketitle

\section{Introduction}

In the current paradigm of hierarchical structure formation, accretion of smaller structures and mergers play a key role in the evolution of galaxies and galaxy clusters. Gas that falls into dark matter halos is heated, mostly via shocks or turbulent dissipation, to the virial temperature of a cluster ($10^7 - 10^8 K$). This results in a fully ionised X-ray emitting intracluster medium (ICM). During the process of cluster formation, mergers and accretion generate hydrodynamic instabilities that create turbulence. Observations and simulations have investigated the properties of turbulent motions in the ICM. In this work, we extend previous theoretical efforts by examining the competing roles of stratification and large-scale infall motions.

There is observational evidence that supports the idea of a turbulent ICM. Examples of observed features used to measure ICM turbulence include X-ray surface brightness fluctuations, interpreted as density fluctuations induced by turbulence, or pressure fluctuations obtained from X-ray maps (\citealt{2004A&A...426..387S}, \citealt{2012MNRAS.421..726S}). Another method to infer information about turbulence is the study of emission lines in the X-ray band. \citet{2010MNRAS.402L..11S} placed limits on the turbulent broadening of the emission lines in cool-core clusters. One of the main findings so far for the internal kinematics of the hot gas was obtained observing the cool-core Perseus cluster with the Hitomi satellite (\citealt{2016Natur.535..117H}). Those authors found a 1D velocity dispersion $\sim 160 \rm ~ km ~ s^{-1}$, on a scale of $L \sim 50$ kpc in the core of the Perseus cluster. 
Similar subsonic turbulent velocities are also suggested by studies of a relatively cold ICM in cool-core clusters 
\citep[e.g.][]{2019A&A...631A..22O, 2019MNRAS.490.3025R}. Indeed, if the cold gas originates by hot ICM cooling, the two phases share the same kinematics (\citealt{2018ApJ...854..167G}).
Turbulence requires a large Reynolds number, which implies a low viscosity. This is also indicated by the evidence of sharp features connected to  Kelvin-Helmholtz instabilities and cold fronts in the ICM (\citealt{2007PhR...443....1M}), which otherwise would be suppressed due to thermal conduction and viscosity (\citealt{2013MNRAS.436.1721R}, \citealt{2018ApJ...868...45W}). Finally, it is possible to exploit the thermal Sunyaev-Zeldovich effect to infer turbulent pressure maps (\citealt{2016MNRAS.463..655K}).

Turbulent motions can indirectly affect other observables of galaxy clusters because they affect non-thermal components, such as magnetic fields and cosmic rays. Combining X-ray and radio observations, \citet{2017ApJ...843L..29E} and \citet{2018MNRAS.478.2927B} found that the brightness fluctuations correlate with diffuse radio emission, suggesting that turbulence might be linked to the acceleration of non-thermal particles. \citet{2006MNRAS.372.1840R} investigated the effect of turbulent diffusion on the metal abundance profile. Moreover, turbulence might play a role in the measurement of cluster masses, as it creates additional kinetic pressure that affects the hydrostatic equilibrium typically assumed to apply to the ICM. In fact, observational and numerical work has confirmed that the ratio between turbulent and thermal energy typically varies between $\sim 10 - 30\%$ \citep[e.g.][]{2020MNRAS.495..864A, 2019A&A...621A..40E, 2019A&A...621A..39E, 2009ApJ...705.1129L}. Since precision cosmology depends on very accurate total cluster mass, turbulent pressure profiles need to be well constrained.
Finally, turbulence has been proposed as a source of heat to prevent development of catastrophic cooling flows  (\citealt{2014Natur.515...85Z}). The most promising scenario there involves turbulent feedback induced by active galactic nuclei. The central gaseous medium is heated by the interaction with buoyantly rising bubbles that are created by jets launched from the central black hole. While such feedback flows can be energetically sufficient, the process by which the  mechanical energy is transferred to the ICM and thermalised is still debated \citep{2002Natur.418..301B, 2010ApJ...710..743D, 2013sncl.confE..88G, 2013AN....334..394G, 2015ApJ...805..112R, 2017MNRAS.464L...1F}.

Fully cosmological, or more idealised simulations have been used to investigate turbulence in galaxy clusters \citep{2005MNRAS.364..753D, 2005ApJ...630L..45K, 2017MNRAS.471.3212W, 2009A&A...504...33V, 2011A&A...529A..17V, 2017MNRAS.464..210V}. \citet{2014ApJ...788L..13Z} and \citet{2014A&A...569A..67G} studied the relation between density and velocity perturbations, \citet{2012A&A...544A.103V} investigated the turbulent diffusion coefficient and \citet{2020MNRAS.495..864A}, \citet{2014ApJ...792...25N}, \citet{2009ApJ...705.1129L} constrained the non-thermal pressure contribution coming from turbulent motions. One of the difficulties in determining ICM turbulent quantities is the disentanglement of bulk and turbulent motions. Several algorithms have been developed for this purpose. For example, \citet{2009ApJ...705.1129L} estimated ICM turbulent velocities as the residual with respect to the velocity field averaged over spherical shells. 
Alternatively, one can estimate the turbulent motions by interpolating the original 3-D velocity field in order to map the local mean field and to identify turbulent velocity fluctuations on scales smaller than the interpolation scale (\citealt{2009A&A...504...33V}, \citeyear{2011A&A...529A..17V}). As another alternative, a multi-scale iterative filtering approach has been implemented in, for example, \citet{2012A&A...544A.103V}, \citet{2020MNRAS.495..864A}, \citet{2021MNRAS.504..510V}. 
In the present work, we apply a simpler approach, that is by using a fixed scale filtering method, based on previous results by our group.

Describing the turbulent nature of the ICM is not trivial. 
Astrophysical turbulent flows are often described in terms of Kolmogorov's theory (\citealt{1941DoSSR..30..301K}), which assumes homogeneity and isotropy.
However, the ICM is a stratified plasma close to hydrostatic equilibrium, where buoyancy may well change the character of the turbulence. In particular, the relation between density and velocity fluctuations, often used to estimate the turbulence strength in the ICM (e.g. \citealt{2014ApJ...788L..13Z}) might vary in the presence of a varying stratification (and forcing mechanism).
The dimensionless Richardson number, given by the square of the ratio of the turbulent eddy turn-over timescale to the buoyancy oscillation timescale (or, equivalently, by the ratio of the buoyancy to inertial forces), is a measure of the importance of the stratification in the turbulence dynamics. It is defined as:

\begin{equation}
    Ri = \frac{N^2_{\rm BV}}{(v_l / l)^2}, 
    \label{eq:Ri}
\end{equation}
where $v_l$ is the characteristic turbulent velocity on a scale $l$, and $N_{\rm BV}$ is the Brunt-V\"ais\"al\"a frequency: 

\begin{equation}
    N_{\rm BV} = \sqrt{- \frac{g}{\gamma}\frac{d \ln}{dr}\left(\frac{P}{\rho^{\gamma}}\right)}.
    \label{eq:NBV}
\end{equation}
Here, $g(r)$ is the gravitational acceleration, $P(r)$ and $\rho(r)$ are the pressure and density, respectively, and $\gamma$ is the adiabatic index of the ICM. Thus, $Ri \ll 1$, implies short turbulent eddy turn-over times compared to buoyancy times, suggesting homogeneous, isotropic turbulence unaffected by density stratification. On the other hand, when $Ri > 1$ the buoyancy times are shorter than the turbulent eddy times. When the buoyancy force becomes dynamically important, it suppresses radial motions and it leads to preferentially azimuthal turbulence (\citealt{2007JFM...585..343B}).

There are previous works investigating turbulence in a stratified ICM-like medium \citep{2014ApJ...788L..13Z, 2014A&A...569A..67G, 2019ApJ...874...42V, 2019MNRAS.487.1072S, 2020MNRAS.493.5838M}. However, they focused on idealised plasma flows, or were limited to the innermost regions of galaxy clusters ($\leq 500 \rm ~kpc$).
Previous observational and numerical work investigated the possibility of inferring the statistics of turbulence in the ICM from gas density and temperature perturbations, which can be recovered from existing X-ray observations   \citep{2016ApJ...818...14A, 2012MNRAS.421.1123C, 2018ApJ...865...53Z}. However, due to the finite sensitivity of X-ray telescopes, this technique is still limited by the large number of photons required to sample large enough cluster scales, with sufficient spatial detail. In order to further compare the results with observational data, past analyses of simulations were restricted to the virial region of the cluster. Moreover, idealised simulations are also bound to miss substructures and filaments, which can bias the analysis compared to the statistics derived from more realistic, dynamically formed clusters.

In this paper we explore the character of the turbulence in realistic, stratified ICM generated in high-resolution, cosmological simulations of galaxy clusters. We mainly focus on the relation between density and velocity fluctuations, their dependence on the Richardson number (see \citealt{2020MNRAS.493.5838M}), and the anisotropy of the turbulent velocity field.

This paper is structured as follows: in Section \ref{sec:methods} we describe the simulated cluster sample and the numerical method used to estimate the relevant variables of a turbulent ICM; in Section \ref{sec:results} we discuss the outcomes of the analysis, looking also for a comparison with previous work; in Section~\ref{sec:conclusion} we summarise our findings, highlighting some observational implications.

\section{Methods}
\label{sec:methods}

\subsection{The Itasca Simulated Cluster sample}

We analysed a sample of galaxy clusters from the 'Itasca Clusters' set of simulations. Each simulation was carried out at high spatial resolution using ENZO, a (parabolic) cosmological numerical code for magneto-hydrodynamics (\citealt{2014ApJS..211...19B}).  A key feature of the code is its Adaptive Mesh Refinement (AMR) capability, which enables a large spatial and temporal dynamical range.

Here we applied a pre-determined set of nested grids in order to apply a constant, comoving grid resolution. Starting from the root grid, which covers the entire volume with a coarse, uniform grid, we placed finer grids as soon as interesting regions started to evolve. In the Itasca suite, we just forced the code to refine the $100 \%$ of the innermost zoom-in region, up to the highest allowed AMR level. 
If the root grid spacing was $\Delta x$, then the spacing of a refined patch at level $l$ was $\Delta x/r^l$, where r is the integer refinement factor.  The simulations presented here employed a purely hydrodynamical  fluid solver based on the Godunov Piecewise Parabolic Method (PPM) scheme. (\citealt{1984JCoPh..54..174C}). 

The cosmological Itasca simulations assume a WMAP7 $\Lambda$CDM cosmology (\citealt{2011ApJS..192...18K}), with $\Omega_B= 0.0445$, $\Omega_{\rm DM}=  0.2265$, $\Omega_{\Lambda}=  0.728$,  Hubble  parameter $h=  0.702$, $\sigma_8=  0.8$ and a primordial index of $n= 0.961$. All runs were non-radiative and did not include feedback from star-forming regions or active galactic nuclei. Even though the  effects  of  non-gravitational heating are small on the $\gg$ 100 kpc scale compared to the heat caused by mergers (\citealt{2019ApJ...874...42V}), the combination of cooling and feedback leads to a higher number of substructures with different temperatures and  densities could create small-scale turbulent motions. In this sense, such structures are absent in our simulations and their presence in the observed galaxy clusters might alter gas density and temperature estimates. Our simulations do not include magnetic fields even though the interplay between turbulence and dynamo amplification is well-established in galaxy clusters \citep[e.g.][]{2012PhRvE..85b6303S,review_dynamo,va18mhd,dom19}. However, the magnetic pressure is small, considering that the typical plasma $\beta$ parameter (i.e. the ratio between thermal and magnetic pressure, $\beta$) is very large, $\beta \sim 10^2-10^3$. This means that we do not expect any relevant modification in the density or velocity structure of the ICM by the (neglected) presence of magnetic fields, except on very small scales ($\leq 30 \rm ~kpc$) where the tension of amplified magnetic field lines is significant \citep[][]{va18mhd}. Nevertheless, the gas dynamics on these small scales is not our primary concern and we proceed with a purely hydrodynamical approach.  

In brief, the Itasca simulations were run according to the following procedure. Firstly, independent cosmological boxes were simulated in order to select the most massive objects (clusters) in the volume. Initial conditions were generated separately for each simulation at redshift $z = 30$. The spatial resolution, on the coarse grid, was $L_0= 110 \rm ~ h ~ kpc^{-1} \approx 157$ ~ kpc and the dark matter (DM) mass resolution was $m_{\rm DM} = 8.96 \cdot 10^7 \rm ~ M_{\odot}$. A second more refined grid was created, centred on the selected cluster formation region, which covered the innermost $\approx$ 31 Mpc. In this case, the comoving spatial resolution reached $55h$ kpc$^{-1} \approx 78.4$ kpc and the DM mass resolution was $m_{\rm DM}= 1.12 \cdot 10^7 \rm ~ M_{\odot}$. Finally, inside the central $(L_0/10)^3 \approx 6.3$ Mpc$^3$ volume of each box, a fixed further refinement was enforced, meaning that $100 \%$ of cells were refined. This volume  was large enough to include the virial radii of most of the clusters, while the spatial resolution was increased to $13.8 \rm ~ h ~ kpc^{-1} \approx 20 kpc$.

The full Itasca cluster sample consists of 20 galaxy clusters and has been used to investigate the dynamic and thermal properties of cluster ICMs. In this work, we used 8 Itasca clusters at $z = 0$ in the mass range $0.54 \cdot 10^{14} \rm ~ M_{\odot} < M_{200} < 3.32 \cdot 10^{14} \rm ~ M_{\odot}$.
For further information on the Itasca cluster sample, we refer to \citet{2017MNRAS.464..210V}, \citet{2017MNRAS.471.3212W}, \citet{2020MNRAS.495..864A}.

\subsection{Filtering method and numerical analysis}
\label{sec:filtering}

In order to disentangle bulk from turbulent motions, we used a tailored filtering method. \citet{2012A&A...544A.103V}, \citet{2017MNRAS.464..210V} have shown that galaxy clusters might present a distribution of turbulent injection scales, since ICM turbulence can be driven by many processes (e.g AGN feedback or accreted clumps). However, \citet{2018MNRAS.481L.120V} utilising the Itasca clusters found that typical scales from turbulence injected by accretion, are in the range $\sim$ 200 - 400 kpc. For this reason, we choose $L = 300$ kpc as a filtering scale in the following analysis. Nevertheless, we tested the robustness of our analysis by using other filtering scales.

Our filtering method estimates the turbulent fluctuation ($\delta v$, $\delta \rho$) as follows:

\begin{equation}
    \delta v = v - \langle v \rangle_L,
    \label{eq:v_filtering}
\end{equation}

\begin{equation}
    \delta \rho = \rho - \langle \rho \rangle_L,
    \label{eq:d_filtering}
\end{equation}
\begin{figure*}[!t]
   	\centering
   	\includegraphics[width= 0.95\textwidth]{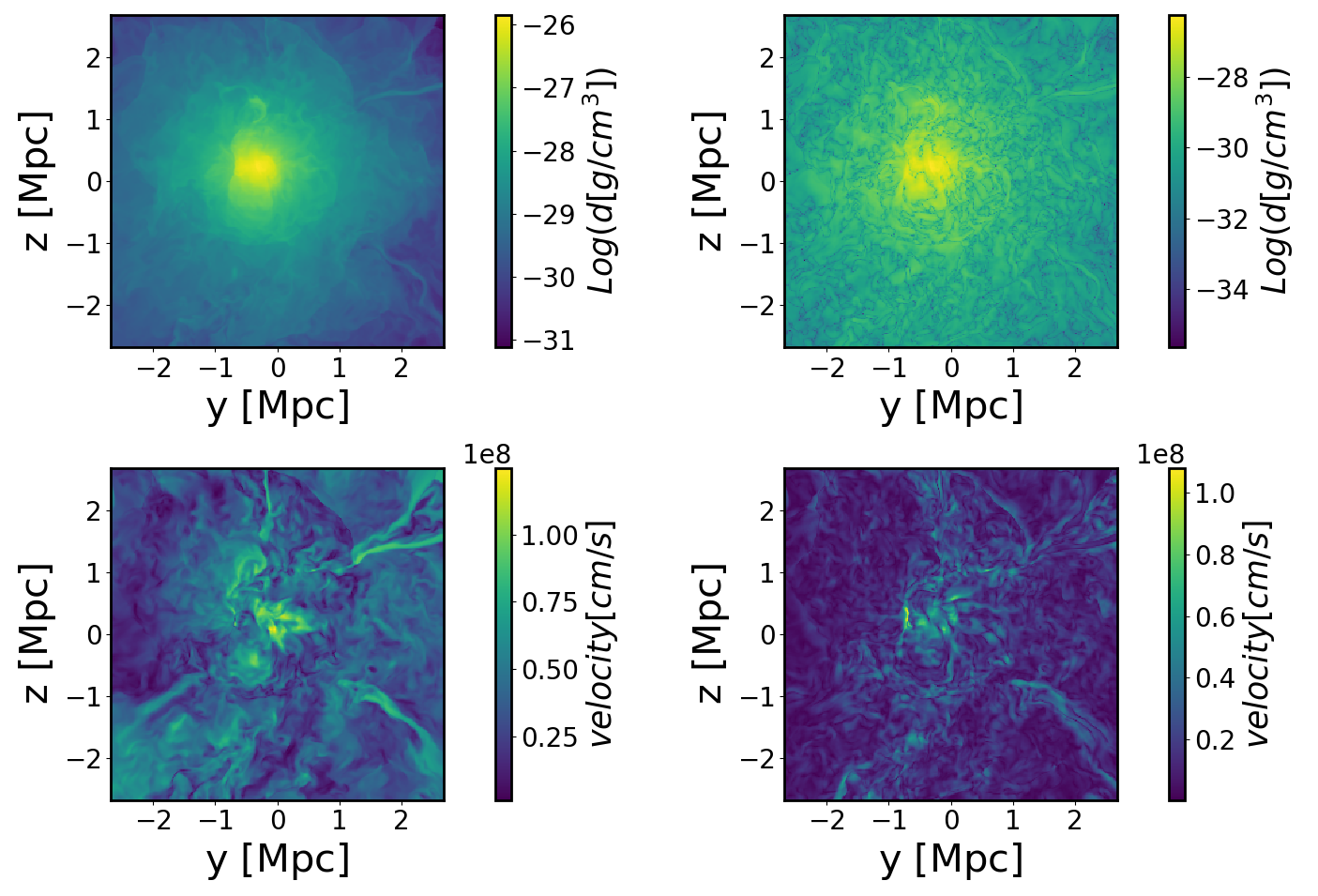}
   	\caption{Slice of the original density and velocity 3D distributions in one cluster (left panels). The right panels show the turbulent density and velocity fluctuations maps, after the application of the fixed-scale filtering method.}
   	\label{fig:turbulent_maps}
\end{figure*}
where $v$ and $\rho$ are the local velocity and density and $\langle \rho \rangle_L$ and $\langle v \rangle_L$ are the means estimated on cubic boxes of (linear) size $L$, centred on the cell to filter. In the case of the velocity, we use a density-weighted mean. Fig.~\ref{fig:turbulent_maps} shows maps of density and velocity fields, before and after we performed this algorithm in a cluster of our sample. We then estimate the density contrast,  $\delta \rho/ \langle \rho \rangle_L$ and the normalised velocity fluctuation, $\delta v / c_s$, where $c_s$ is the sound speed at a given radius, obtained from the (azimuthally averaged) temperature profile. We investigate the relation between these two quantities to directly compare to the work by \citet{2014ApJ...788L..13Z}.

In the following, we calculate the statistical quantities (for instance the variance of the density contrast) in cubic boxes of linear size 600 kpc, located at different radii and different directions. In this way we map the whole simulation box to assess the spatial variation of the turbulence properties.
We also performed a similar analysis by dividing the main box in spherical shells, with similar but somewhat less clear results. This can be understood by the azimuthally asymmetric ICM dynamics, generated by accretion events along the large scale cosmological filaments. Because of this we only discuss the sub-boxes analysis in the rest of the paper. 


\subsubsection{Density and velocity fluctuations}
\label{sec:fluctuation}

     We adopt as a measure for the typical density fluctuation the standard deviation of the probability distribution function (PDF) of the density contrast $\delta \rho/\langle\rho\rangle $, which we indicate in the following with $\sigma_{\rho}$, computed in every box.
     In order to limit the contamination from shock compression and substructures, we eliminate the cells corresponding to the top 5\% of the density contrast distribution in every box, following earlier work concerned with X-ray modelling \citep[e.g.][]{2013MNRAS.432.3030R}. For testing purposes, we also experimented with other thresholds, obtaining: $\sigma^2_{\rho,99} = 0.26, \sigma^2_{\rho,95} = 0.22 $ and $\sigma^2_{\rho,90} = 0.20$, which refer to thresholds of 99\%, 95\%, 90\%, respectively.In a similar way, the mean normalised velocity fluctuation, $\sigma_v$, is estimated as the standard deviation of $\delta v/c_s$, defined in the previous section, using the same cells used to calculate $\sigma_\rho$.
     
\begin{figure}
        \centering
        \includegraphics[width= 0.5\textwidth, height=0.3\textwidth]{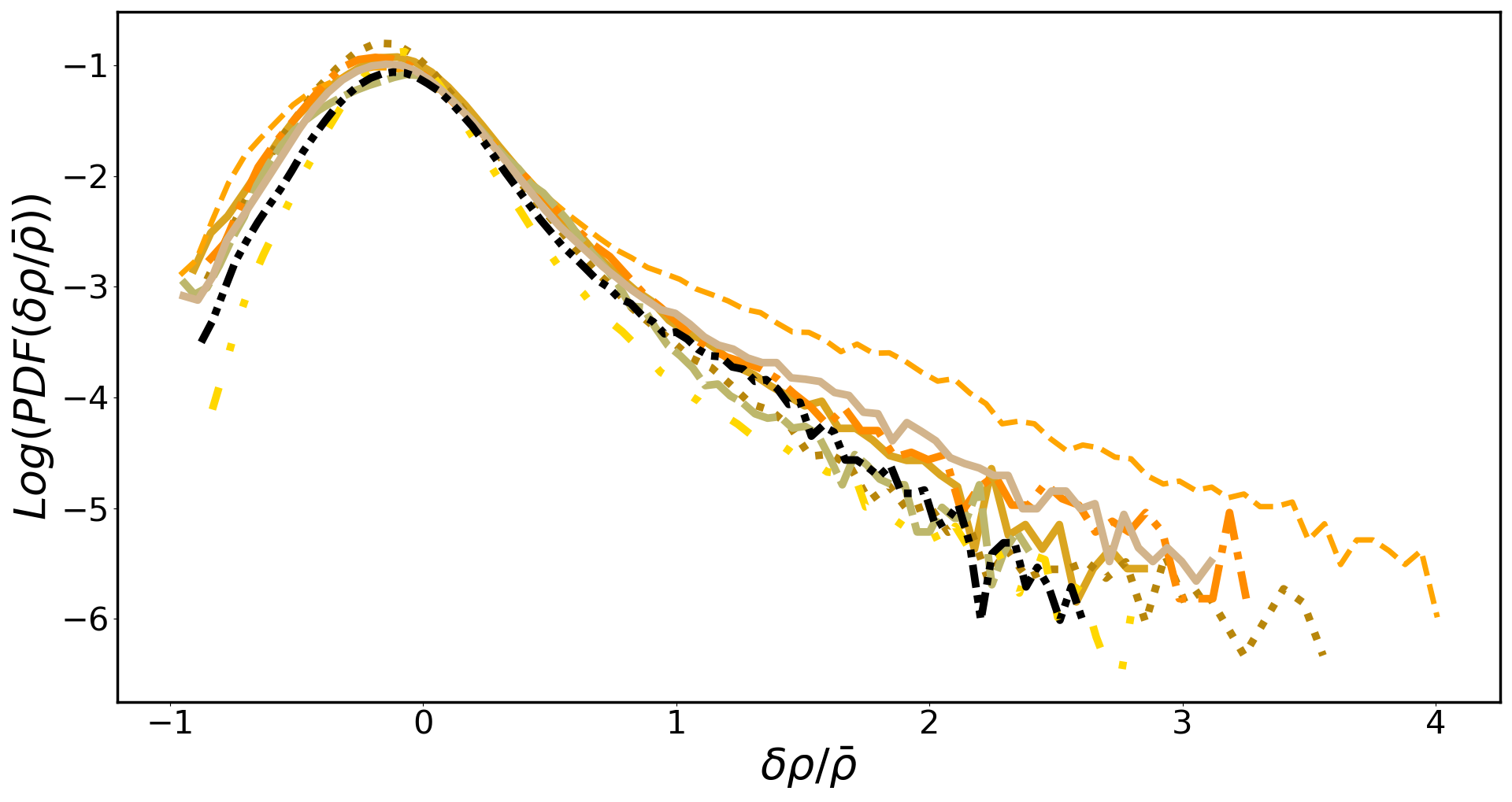}\quad\includegraphics[width= 0.5\textwidth, height=0.3\textwidth]{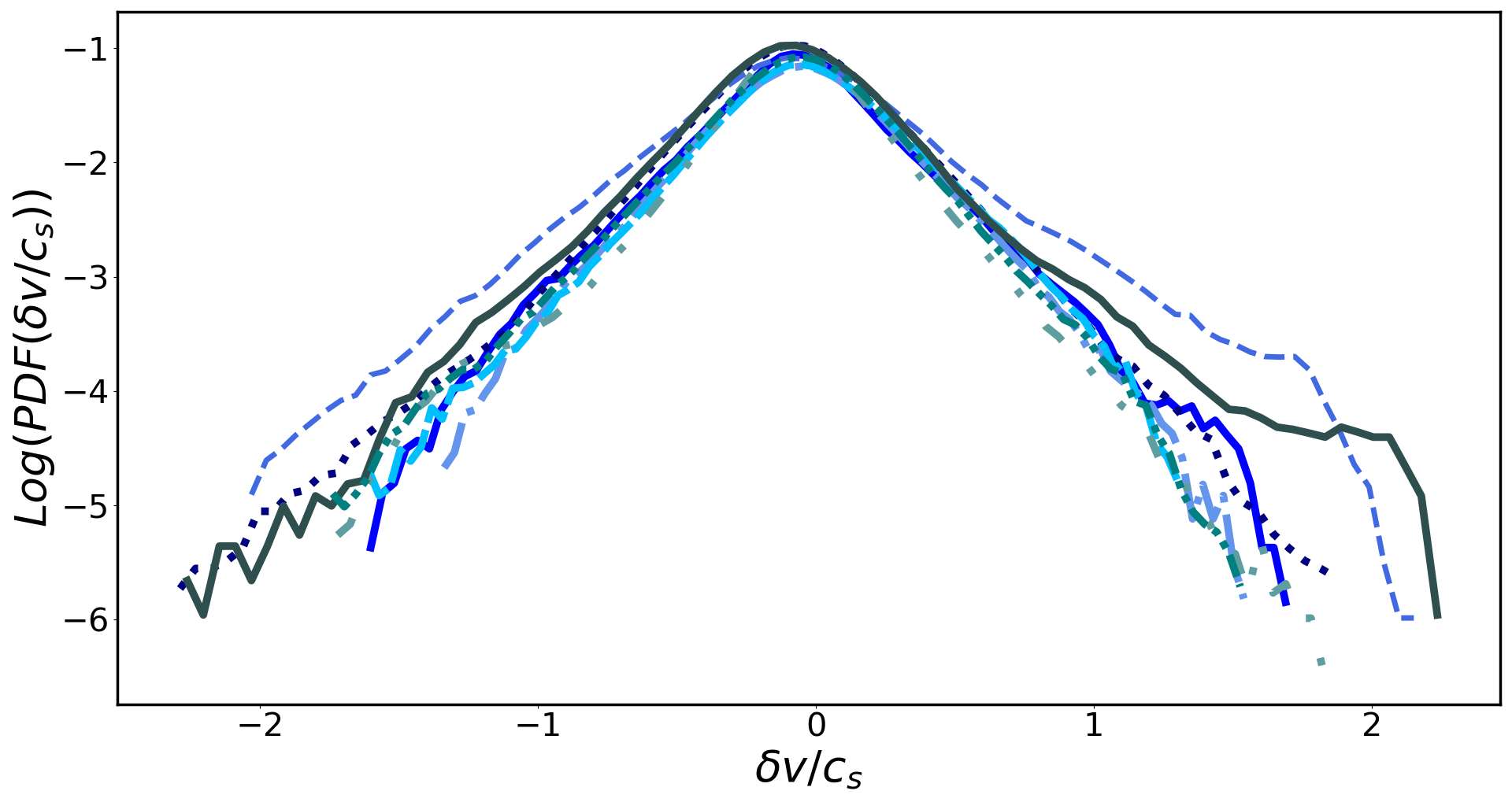}\quad\includegraphics[width= 0.5\textwidth, height=0.3\textwidth]{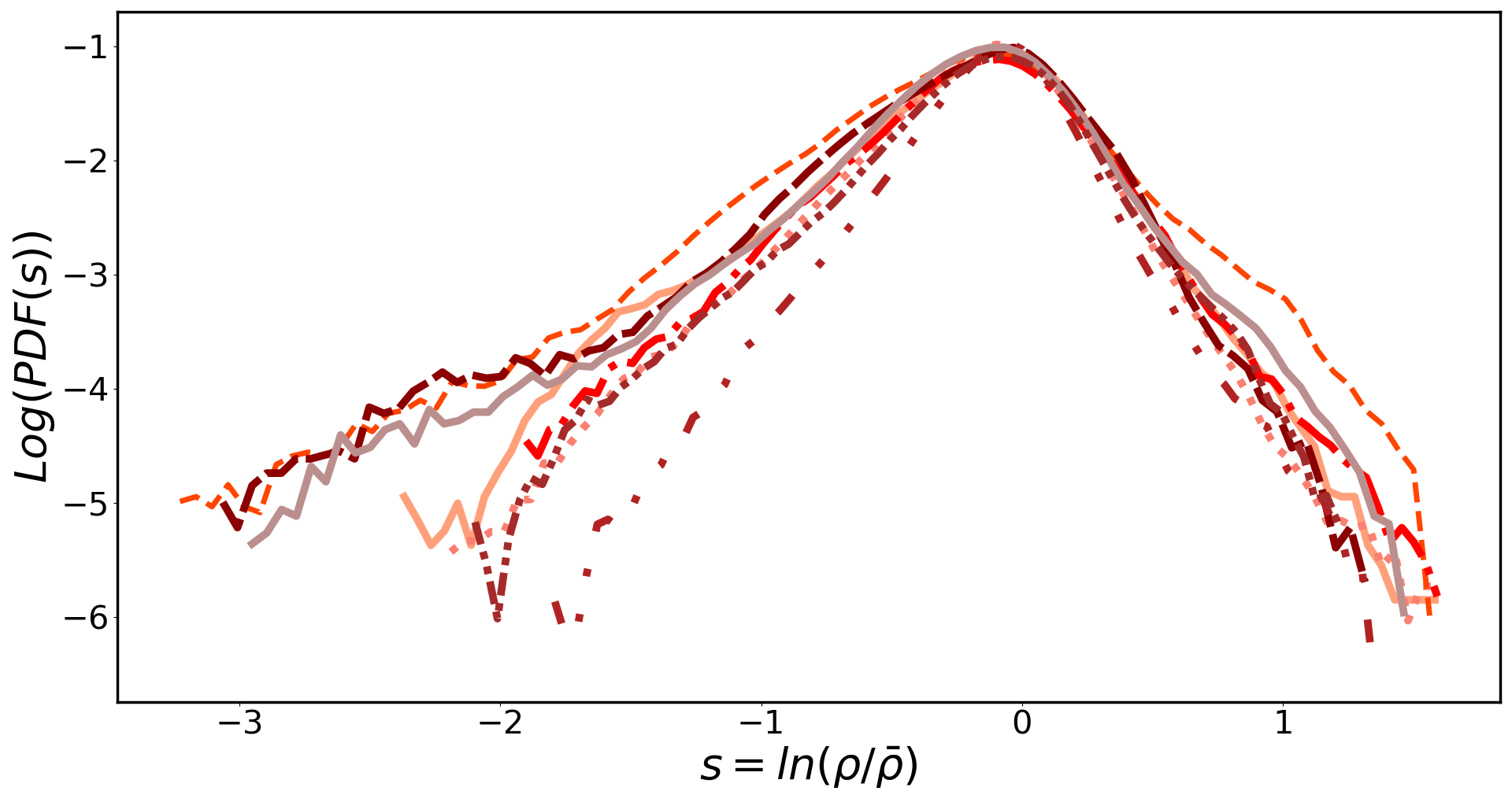}
        \caption {Probability distribution function of the gas density contrast, logarithmic gas density fluctuations, and velocity fluctuations. The colours correspond to different clusters, at z = 0. All statistics refer to within the virial radius.}
        \label{fig:pdf}
  \end{figure}

     Finally, to compare our results on stratified turbulence with previous work, we also consider the logarithmic density fluctuations, $s = \ln(\rho/ \langle \rho \rangle_L)$ and calculate the standard deviation in every box, $\sigma_s$. Fig.~\ref{fig:pdf} shows, for each cluster of the sample, the probability distribution function of the variables we use in this work: the gas density contrast, $\delta \rho / \bar{\rho}$, the logarithmic gas density fluctuations, $ s = ln( \rho / \bar{\rho})$, and the velocity contrast $\delta v / c_s$. The PDFs are obtained from the fluctuations within the virial radius (defined as, $R_{200}$, i.e. the radius enclosing a $200$ overdensity with respect to the cosmic critical density).
     
     \subsubsection{Richardson number}
     \label{sec:Ri}
     
     The Richardson number, $Ri$, is central to our study, but, in contrast to some previous work in which the authors considered $Ri$ as an input parameter of the simulation (\citealt{2020MNRAS.493.5838M}, \citealt{2019MNRAS.487.1072S}), we had to compute it a-posteriori following Eq.~\ref{eq:Ri}. We obtained the radial profile of the Richardson number in 600 kpc wide boxes, starting from the gravitational acceleration and gas pressure and density radial profile in the box that we used to estimate the Brunt-Väisala frequency , $N_{\rm BV}$ (Eq.~\ref{eq:NBV}). We also evaluated the turbulent velocity profile in the box, computing a root mean square after the excision of the densest density fluctuations (see Sec.~ \ref{sec:fluctuation}).  For the sake of simplicity, we fixed the spatial scale in Eq.~\ref{eq:Ri} as the size $L$ of the filtering scale (see Sec.~\ref{sec:filtering}). In order to characterise each box with its Richardson number, we averaged the $Ri$ radial profile.
     
\begin{figure}
        \centering
        \includegraphics[width= 0.5\textwidth, height=0.3\textwidth]{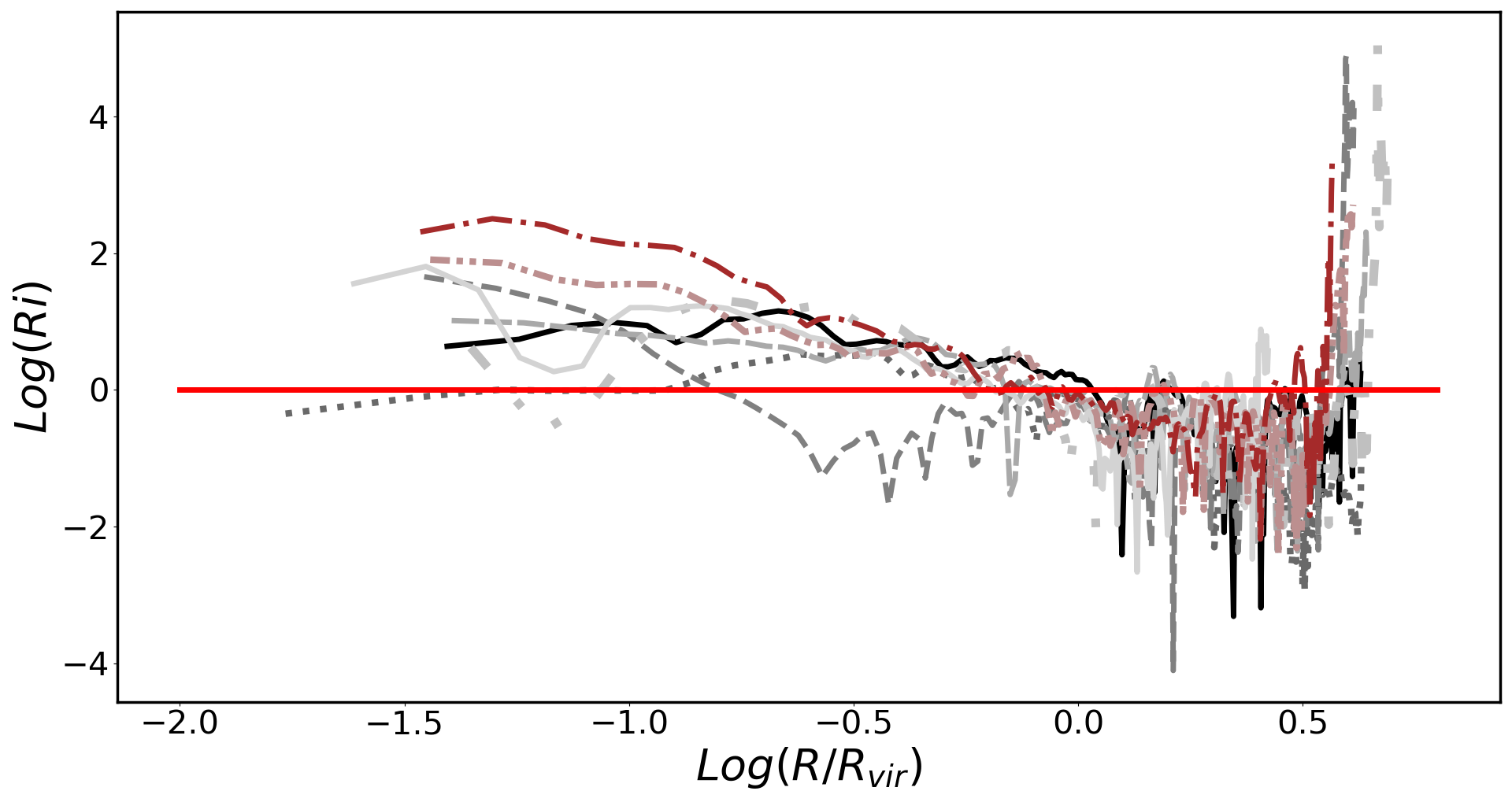}\quad\includegraphics[width= 0.5\textwidth, height=0.3\textwidth]{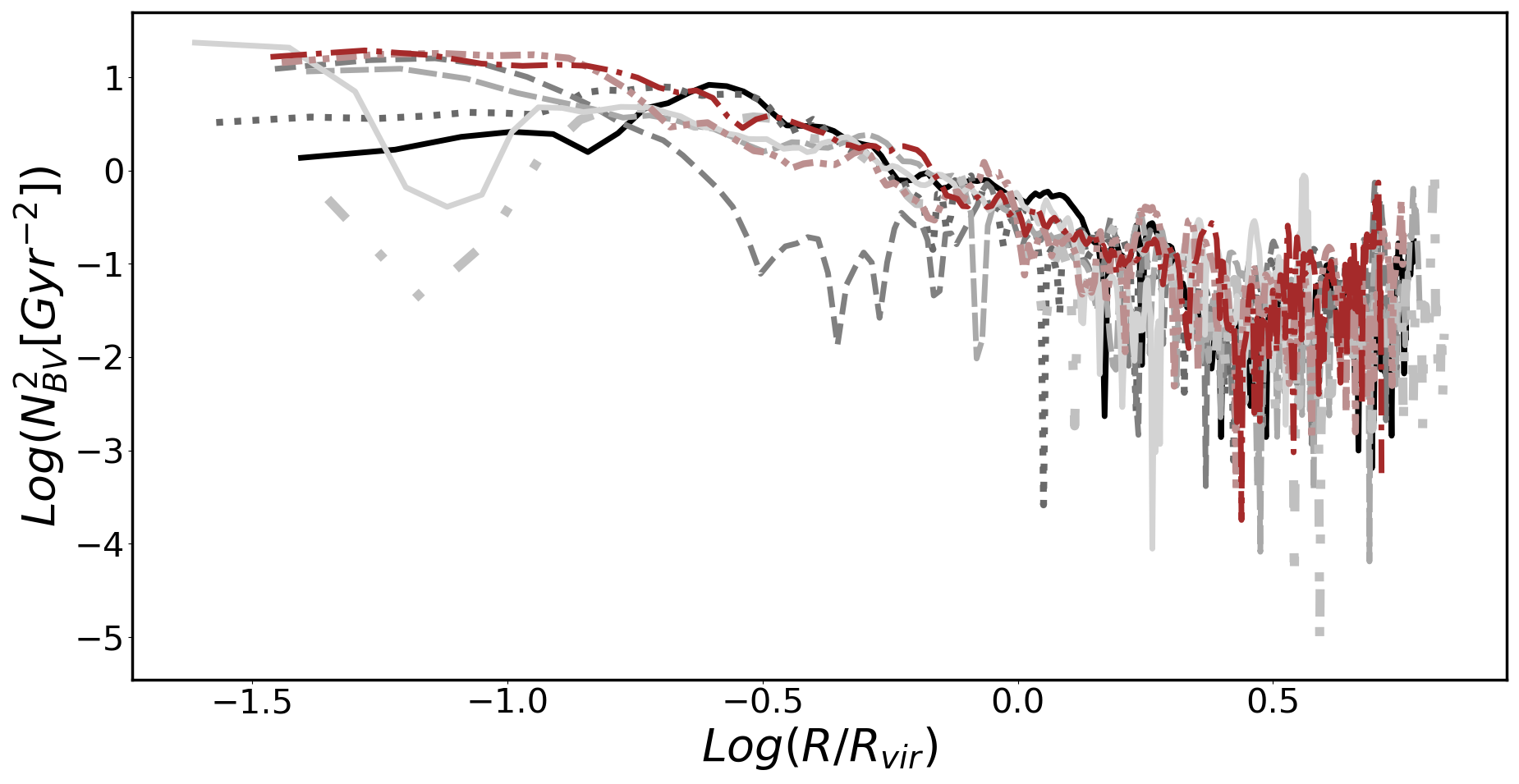}\quad\includegraphics[width= 0.5\textwidth, height=0.3\textwidth]{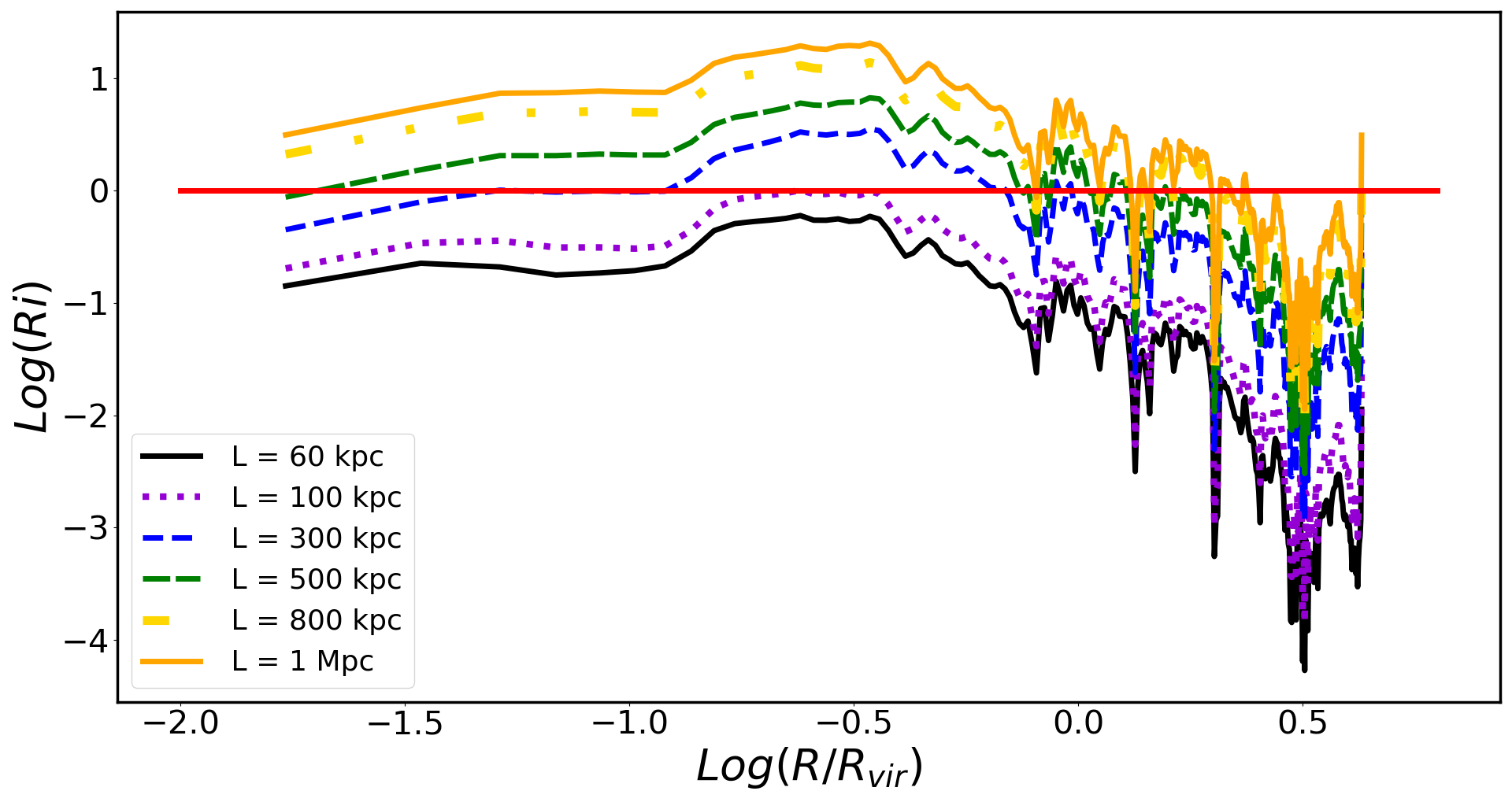}
        \caption{Top panel: radial profile of the Richardson number in our clusters.
        Middle panel: radial profile of the Brunt-Väisala frequency for each cluster of the sample. 
        Lower panel: radial profile of the Richardson number considering different filtering scales $L$, for a single cluster of our sample. The horizontal red line separates the turbulence-dominated ($Ri$ < 1) and buoyancy-dominated ($Ri$ >1) regimes.}
        \label{fig:Ri_profile}
     \end{figure}

     Furthermore, we obtained the radial profile of the Richardson number over the whole volume of each galaxy cluster. The top panel of Fig.~\ref{fig:Ri_profile} shows the Richardson number profile in our cluster sample for a filtering scale $L$ = 300 kpc (top panel), together with the Brunt-Väisala frequency profile (middle panel). In all clusters, except one, buoyancy is always found to be dominant ($Ri \gg 1$) within the virial radius, while the turbulent cascade overcomes gravity only in the outskirts. Moreover, we find that the Richardson number has a strong dependence on the filtering scale $L$ (Fig.~\ref{fig:Ri_profile}, lower panel). It nearly follows the Kolmogorov scaling $Ri \propto L^{4/3}$. Consequently, a local estimate of $Ri$ must always be referred to a turbulent scales.
    
    As noted above, in the remainder of this work we restrict ourselves to the fixed filtering scale of $L=300 \rm ~kpc$; this is motivated by our  previous works on the Itasca dataset.
    For example, in \citet{2017MNRAS.464..210V} and \citet{2020MNRAS.495..864A} we measured, with an iterative multi-scale filtering algorithm, that the kinetic-energy weighted distribution of cells in the volume of our clusters have a local turbulent outer scale $\leq 400 \rm ~kpc$, with little dependence on the dynamical state of the host cluster. Based on these earlier studies, we can conclude that $L \geq 500 \rm ~kpc$ appears to be too large based on the actual kinematic analysis of these systems, while $L \leq 200 \rm ~kpc$ scales are too small to capture the total turbulent kinetic energy budget developed in the ICM of these simulated systems.

     In order to quantify the degree of anisotropy, possibly caused by the buoyancy, we compute the parameter $\beta$,
     
     \begin{equation}
         \beta = 1 - \frac{v^2_t}{2 v^2_r},
     \end{equation}
     
     where $v_t$ and $v_r$ are the tangential and radial components of the turbulent velocity distribution (that is, the r.m.s. values computed in 600 kpc wide boxes, with the filtering scale again set to $L=300$ kpc), respectively. The anisotropy parameter can span the range $-\infty < \beta \leq 1$. Positive values mean that the dynamics is dominated by the radial motions, whereas, when $\beta < 0$ the dominant components of the velocity field are those perpendicular to the radial direction. We refer the reader to \citet{2018MNRAS.481L.120V} for a further analysis of the anisotropy parameter in the Itasca cluster sample.

     \subsubsection{Power spectra}
     
     In this section, we compare the density and velocity power spectra at various cluster radii. 
     We measure the power spectra of the ICM velocity field in large boxes, $\approx$ 2.5 Mpc (linear size) wide, located at different radii using a Fast Fourier Transform, with a periodic domain for simplicity \citep[e.g. see][for a discussion]{2011A&A...529A..17V}:
     
   \begin{equation}
   \abs{\textbf{v}(\textbf{k})}^2 = \sqrt{ v_x(\textbf{k})^2 + v_y(\textbf{k})^2 + v_z(\textbf{k})^2 },
   \end{equation}
   where ${v}_i(\textbf{k})$ is the Fourier transform of each component
   
   \begin{equation}
   v_i(\textbf{k}) = \frac{1}{\left (2 \pi \right) ^3} \int_V v_i(\textbf{x}) e^{-2\pi i \textbf{k} \cdot \textbf{x}} \,d\textbf{x}.
   \end{equation}
   
\begin{figure}
        \centering
        \includegraphics[width= 0.5\textwidth]{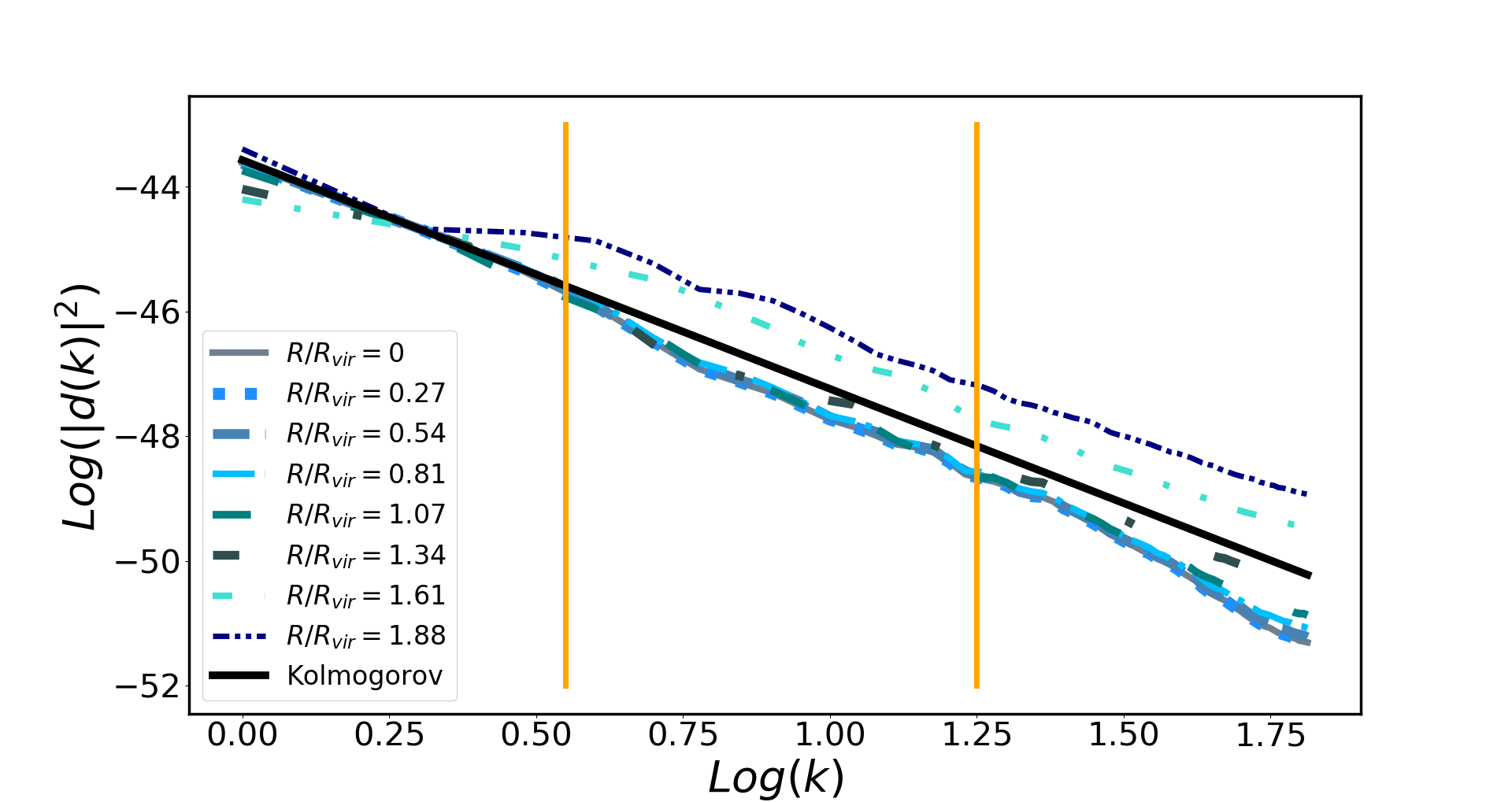}\quad\includegraphics[width= 0.5\textwidth]{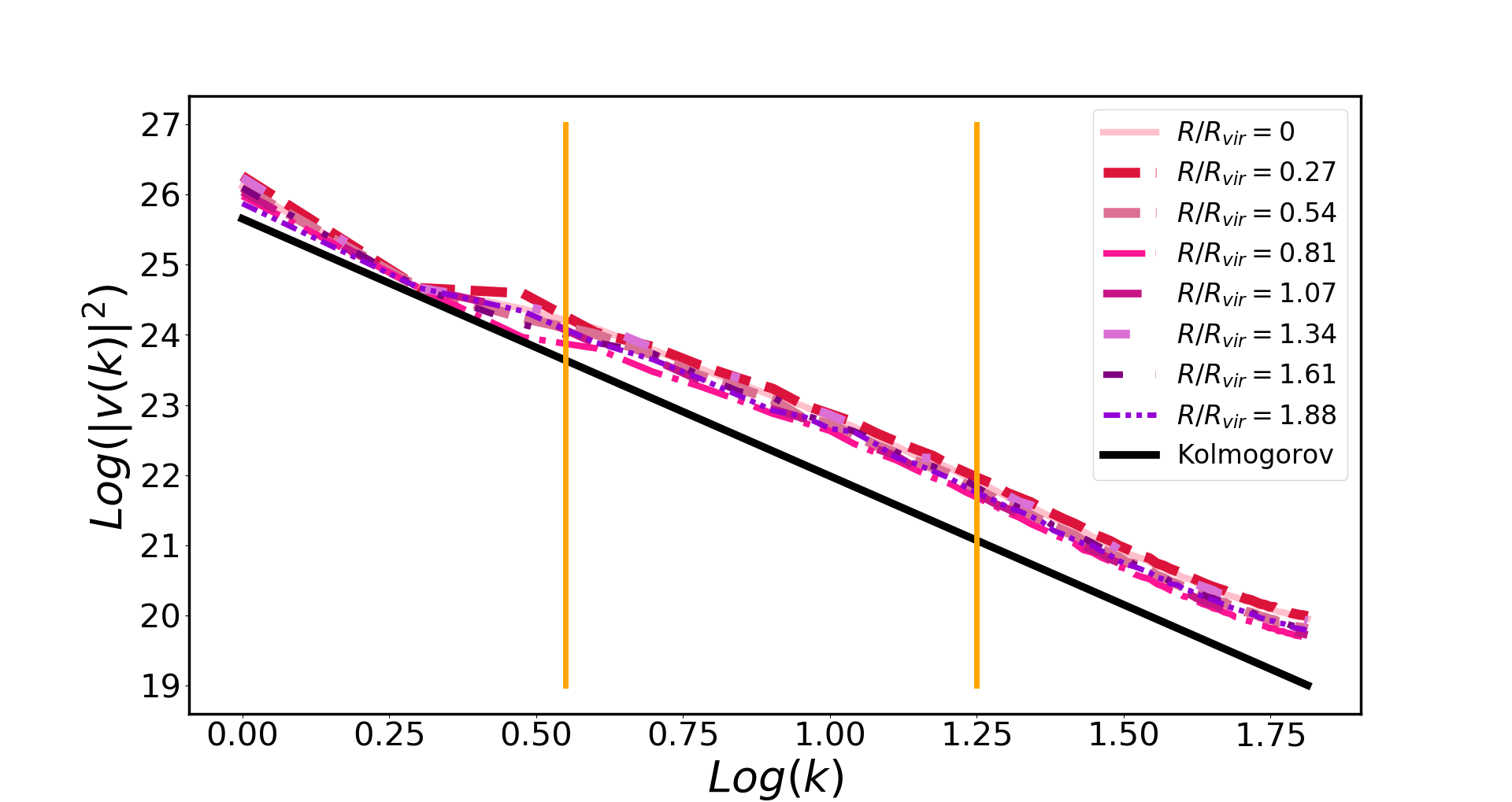}
        \caption {Density and velocity power spectra in a simulated cluster from the Itasca sample, obtained at different radii. The vertical orange bars indicate the selected inertial subrange used for the analysis. The normalisation of density spectra has a radial trend, with larger fluctuations in the outer regions.}
        \label{fig:power_spectra}
  \end{figure}

   The gas density fluctuations, $\abs{\rho (\textbf{k})}^2$, are obtained in a similar way, that is by computing the Fourier transform of the real space density distribution. In this case we did not perform any filtering, since filtering with a fixed spatial scale would just suppress the power on $k \leq 2 \pi/L $ scales, but we select the inertial subrange highlighted with the orange bars (Fig.~\ref{fig:power_spectra}):  150 kpc $\lesssim l \lesssim$ 1 Mpc. The small scale limit is being fixed to avoid measuring spectral shapes in a regime already affected by the numerical dissipation of the PPM hydro scheme. The normalisation of the power spectra in Fig.~\ref{fig:power_spectra} has been rescaled to compare the slopes at various radii. In general, the normalisation of the density spectra increases outside the virial radius, except for the power spectrum computed in the outermost region of the cluster which shows a drop. This increasing trend is likely due to the larger clumping factors in the outskirts \citep{2021arXiv210201096A}.
   
We evaluated the slope of each $k$-bin within the inertial subrange and
we computed an average value for a more robust characterisation of the power spectrum. For the velocity power spectra we evaluated the slope in each bin of each velocity component and then we averaged over the all slope values. Finally, we averaged the slopes of the power spectra performed on boxes at the same radius, computing an inverse-variance weighted mean, to investigate the radial trend of the slope of the power spectrum, compared to the $\propto k^{-11/3}$ trend expected for the Kolmogorov model in three dimensions.
   
\begin{figure}
        \centering
        \includegraphics[width= 0.5\textwidth]{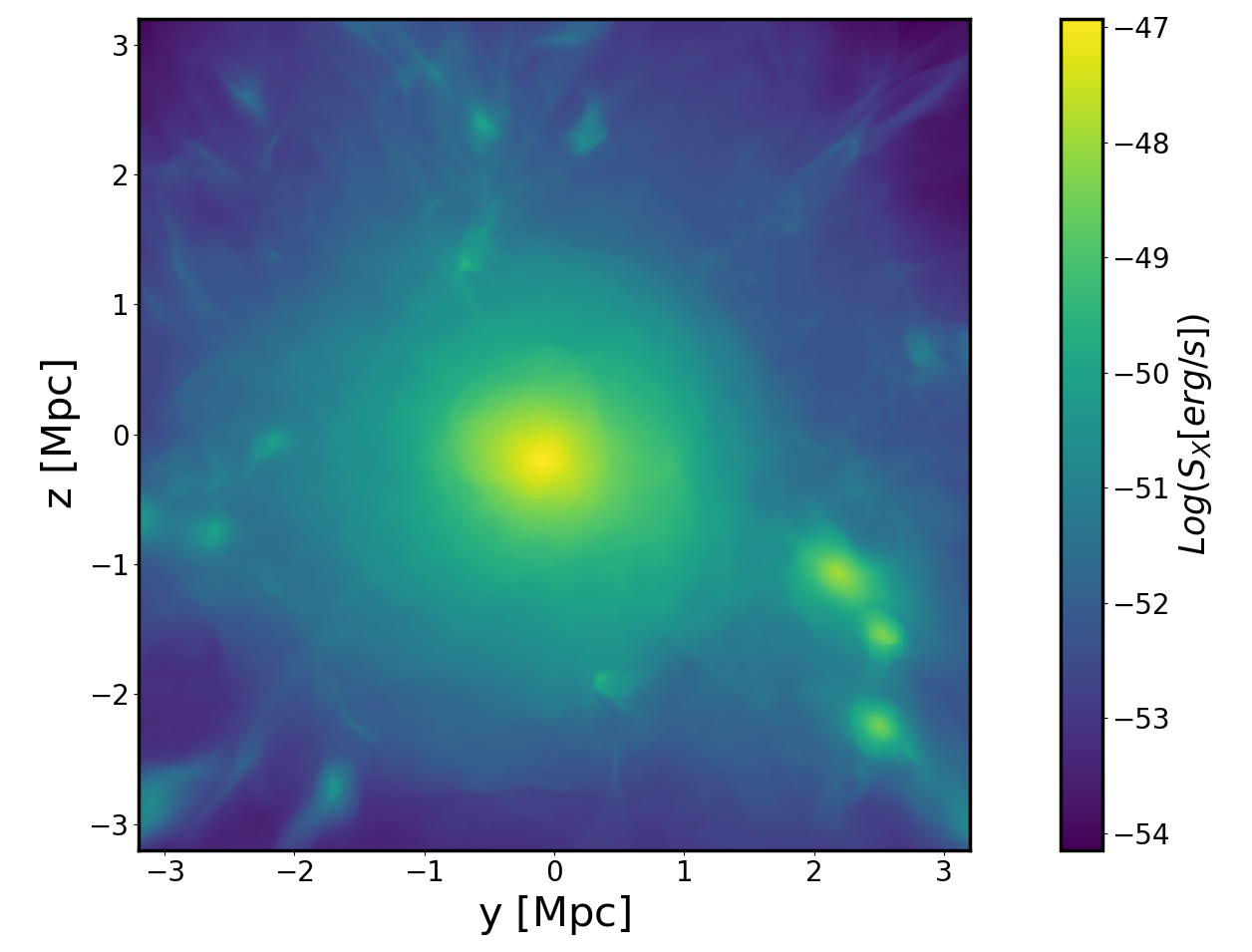}\quad\includegraphics[width= 0.5\textwidth]{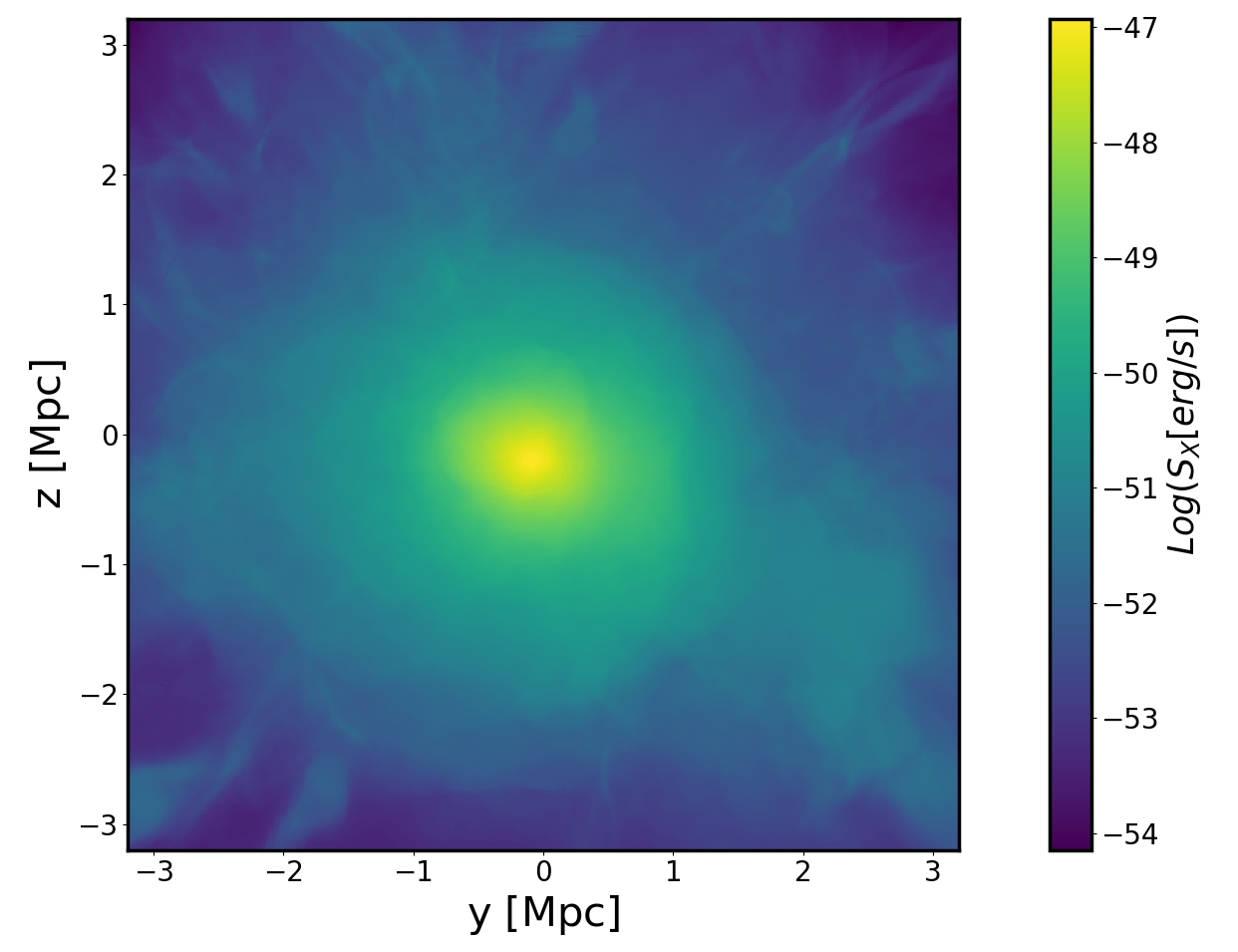}
        \caption{Projected X-ray surface brightness in a cluster of our sample before (top panel) and after (bottom panel) filtering out the clumps.}
        \label{fig:Sx_map}
     \end{figure}

\begin{figure}
        \centering
        \includegraphics[width= 0.5\textwidth]{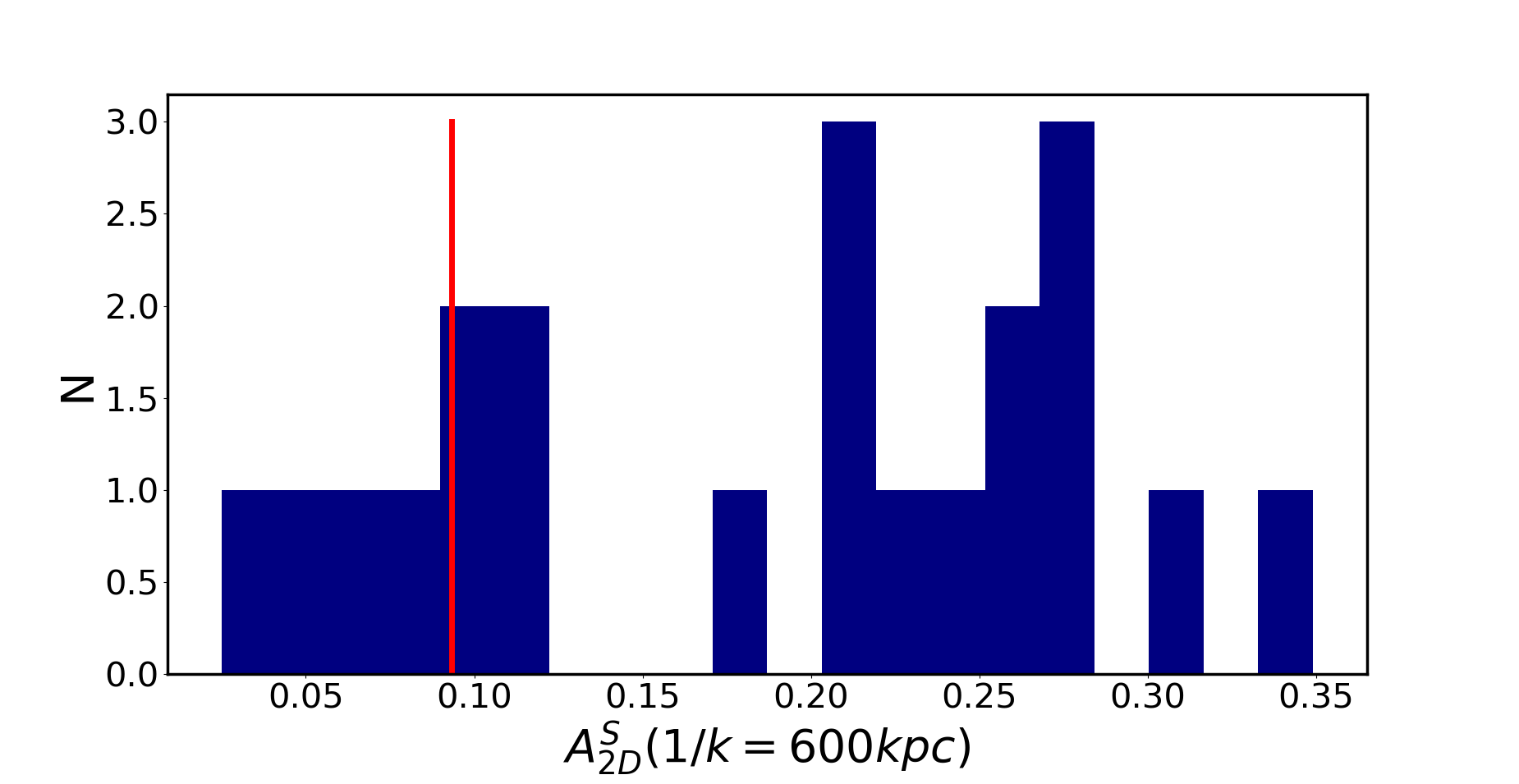}
        \caption{Distribution of the X-ray surface brightness fluctuations, $A_{2D}^S$, in the Itasca cluster sample.}
        \label{fig:Sx_distribution}
     \end{figure}

  As a preliminary check on the reliability of our cluster simulations for this study,  we 
  computed the statistics of gas fluctuations. As previously explained, some spurious features that can affect our numerical analysis might be present in our galaxy clusters (e.g. shocks, clumps or filaments), as well as to some degree also in real observations. In order to excise most of such features not associated with true turbulence-driven fluctuations, for each radial shell we computed  the probability distribution function of density and excluded the 5\% densest cells. In order to have uniformly filled maps, we replaced the original density values of these cells with the median density within the entire radial shell.  This method has been implemented in previous works \citep[see][for an example]{2011MNRAS.413.2305V}. Considering that bremsstrahlung is the dominant radiative mechanism at the typical temperature of plasma in galaxy clusters ($10^7 - 10^8 K$), we estimated the surface brightness by projecting $S_X = n^2 T^{1/2}$ (where $n$ and $T$ are the plasma density and temperature in each cell) along multiple lines of sight. Fig.~\ref{fig:Sx_map} shows the projected X-ray surface brightness in one of our simulated clusters before and after removing the clumps. We can thus estimate the surface brightness fluctuations, dividing the X-ray images by their radial average value within each radial bin. This method is commonly used in observational studies, even though it does not consider any azimuth asymmetries of the ICM. Thus, we obtain the 2D power spectra of X-ray fluctuations on 1~Mpc wide boxes centred on the gas density peak of the cluster. The characteristic amplitude is computed as follows \citet{2012MNRAS.421.1123C}: 
  
  \begin{equation}
      A_{2D}^S(k) = \sqrt{2 \pi P_{2D}(k) k^2},
      \label{eq:amplitude}
  \end{equation}
  where $P_{2D}(k)$ is the 2D power spectrum and $k$ is the wavenumber in $kpc^{-1}$ units. The distribution of surface brightness fluctuations in our cluster sample is shown in Fig.~\ref{fig:Sx_distribution}. \citet{2017ApJ...843L..29E} computed the distribution of the amplitude in surface brightness fluctuations for 51 galaxy clusters, and looked for the possible correlation between larger fluctuations and the presence of non-thermal radio emission. The results can be used to investigate the connection between turbulence and the re-acceleration of relativistic electrons via Fermi II mechanism \citep[e.g.][and references therein]{bj14}. The red line in Fig.~\ref{fig:Sx_distribution} represents the mean value of the X-ray surface brightness fluctuations in \citeauthor{2017ApJ...843L..29E} galaxy cluster sample, which is 0.9. We observe that our fluctuations are by a factor $\sim 2$ larger.\newline
  We should mention that our statistics are not affected by observational limitations. The surface-brightness fluctuations can be reconstructed for the entire cluster volume with equal fidelity and ignoring such limiting factors as the number of emitted X-ray photons. Real observations, however, are affected by the uneven effective spatial resolution, and by the fixed sensitivity to X-ray brightness fluctuations. Moreover, it is not straightforward to implement exactly the same procedure for the masking of point-like X-ray sources in observations, which can bias the observed amplitude low by removing a part of gas density enhancement associated with the surroundings of excised galaxies and AGN. 
  

   \subsubsection{Structure function}
  
\begin{figure}[h!]
        \centering
        \includegraphics[width= 0.5\textwidth]{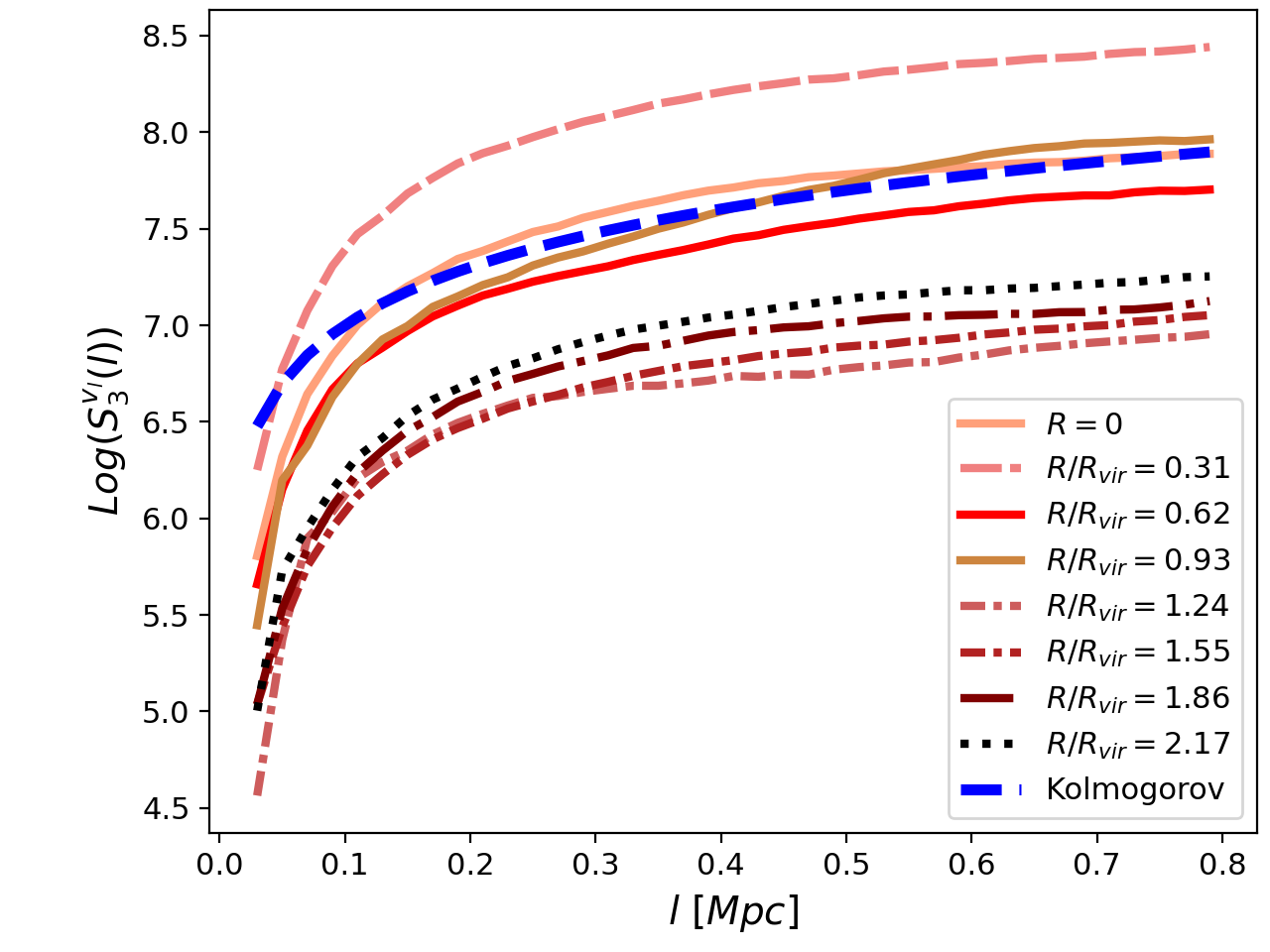}\quad\includegraphics[width= 0.5\textwidth]{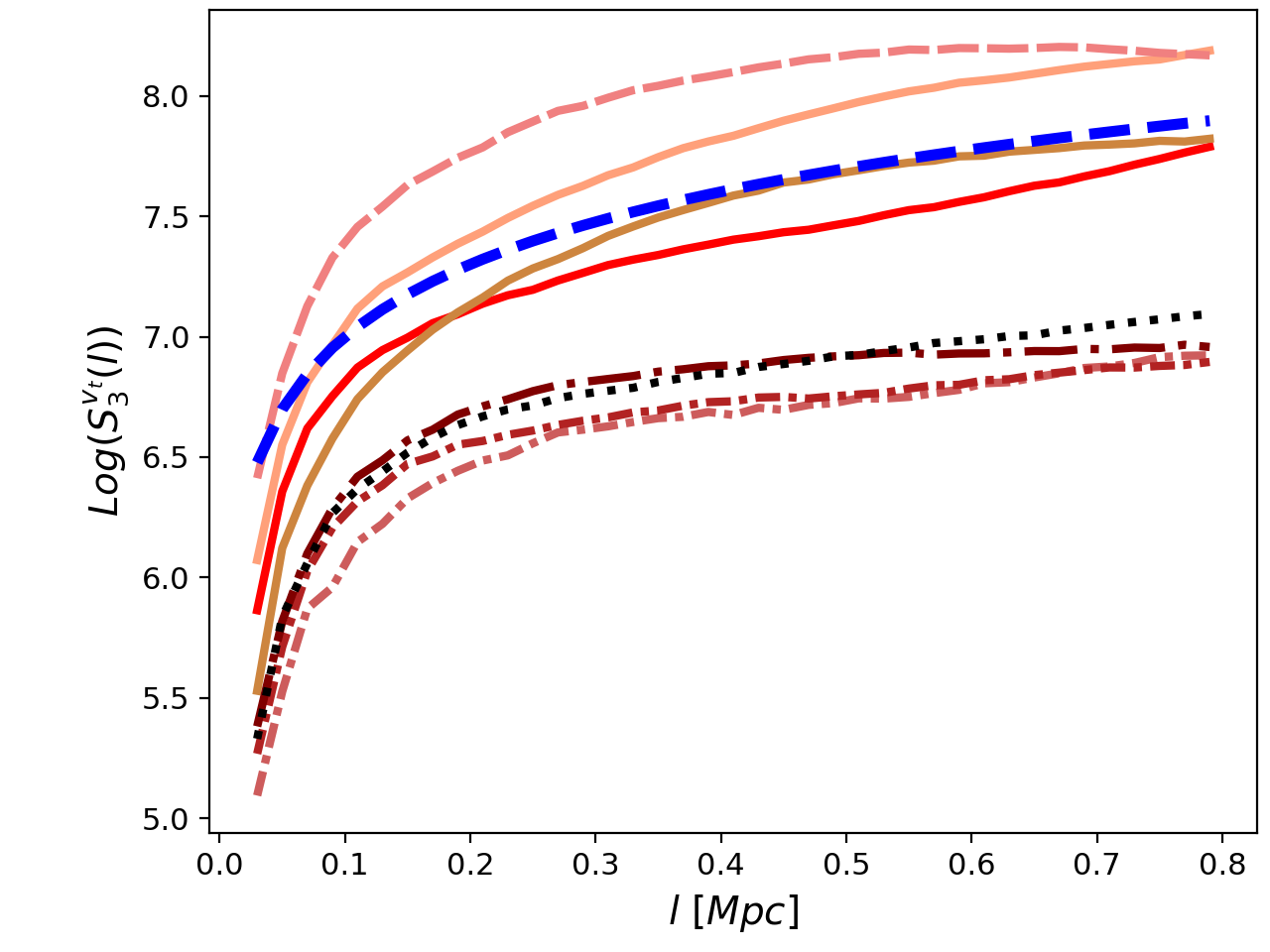}\quad\includegraphics[width= 0.5\textwidth]{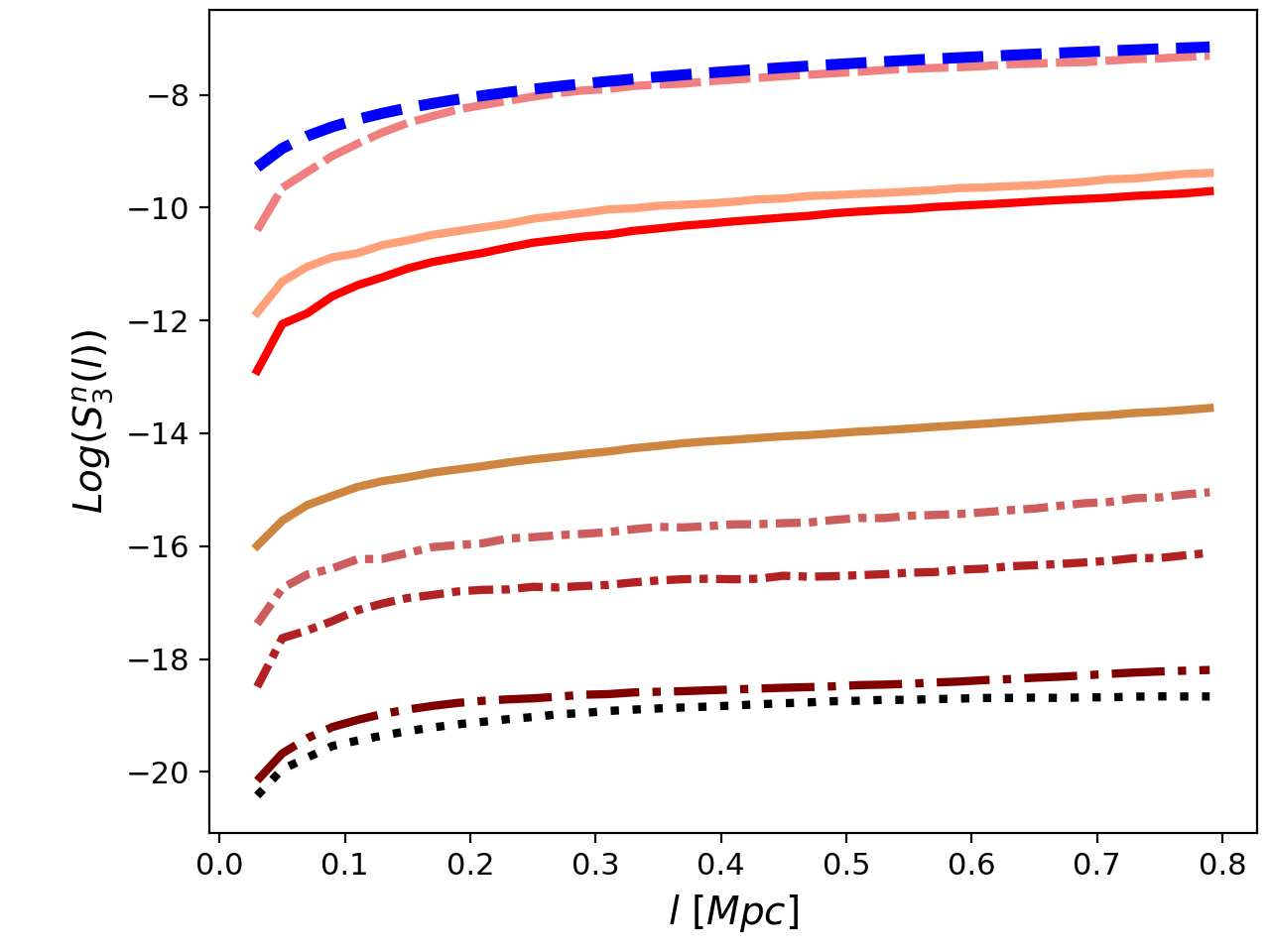}
        \caption{Structure function of the tangential (upper panel), transverse (middle panel) and numerical density (lower panel). The slope is consistent with Kolmorov's theory in the range between 100 - 800 kpc, while it has a steeper behaviour at the smallest scales ($l < 100 \rm ~ kpc$). The normalisation increases with cluster-centric distance, especially for the density structure function due to the stratification of the ICM.}
        \label{fig:SF}
     \end{figure}

  For the sake of completeness, we also computed the third-order structure function of the velocity and density for the same reference galaxy cluster used above for the power spectra analysis, as usually this is more commonly done in study of other turbulent astrophysical environments \citep[e.g.][]{2007ApJ...665..416K}.
  Fig.~\ref{fig:SF} shows the third-order structure function of the velocity component tangential (upper panel) and longitudinal (central panel) to the separation, $l$, and of the particle density distribution (lower panel). The analysis was carried out on 800 kpc wide boxes, located at multiple distances from the centre of the cluster. We do not use any previously applied filtering method because the influence of the laminar flow is dominant mainly at the largest scales ($\gtrsim 1 \rm ~ Mpc$). The maximum separation distance explored here is $l \sim \rm ~800 kpc$. We refer the reader to \citet{2011A&A...529A..17V}, (\citeyear{2017MNRAS.464..210V}) for further details on how the analysis was performed. Our analysis shows that, overall, both the gas density and velocity third-order structure functions have a Kolmogorov-like scaling, $S_3(l) \propto l$, within the separation range 100-800 kpc, similarly to the velocity power spectra. As in previous work \citep[][]{2011A&A...529A..17V}, the steepening at the smallest separations is mostly driven by the increasing dissipation of the PPM scheme, which dampens fluctuations faster than in the Kolmogorov model. 
  The normalisation of the velocity structure function increases from the outskirts to the centre of the cluster, with some variance related to the presence of substructures. The normalisation of the density structure function, on the other hand, decreases in a more steady way towards the periphery, following the drop of the cluster gas density profile (see Fig.~\ref{fig:turbulent_maps}), which contributes to $S_3(l)$ with the density cube. 
  
  Recent studies investigated the structure function in galaxy clusters \citep{2020ApJ...889L...1L, 2021MNRAS.504..898W, 2021arXiv210901771M}. Overall, the structure functions have a Kolmogorov slope ($\propto l^{1/3}$ for first-order functions), or steeper. In those studies the turbulence is driven by the black hole energy output and the observed scales range between 1 and 100 kpc, which cannot be probed by our simulations. On the other hand, our simulations are more suitable to capture the development of turbulence injected by mass accretion on $\sim 0.1-1 \rm ~ Mpc$ scales, and the complementary view of the structure function confirms that the dynamics of gas motions are reasonably close to a Kolmogorov model.
  
  \subsection{Testing our method in idealised simulations}
  \label{sec:test}
  
  Next, we validate our method against more idealised simulations. As previously discussed, due to the different nature of the simulations, it is not feasible to apply the very same algorithm that \citealt{2020MNRAS.493.5838M}, \citeyear{2021MNRAS.500.5072M} used in their studies to our cluster sample. For instance, in \citeauthor{2020MNRAS.493.5838M} and \citeauthor{2019MNRAS.487.1072S}, the Mach number and the Richardson number were parameters of the simulation. However, in cosmological clusters these quantities cannot be estimated a priori. Furthermore, the algorithm to evaluate the density fluctuations must be revisited as well, due to the presence of self-gravitating structures inherent in forming clusters, including the Itasca clusters (see Sec.~\ref{sec:filtering}).
  Therefore, we applied our analysis to data from the simulations by \citeauthor{2019MNRAS.487.1072S}. In their work, they used the FLASH code \citep{2000ApJS..131..273F} to simulate 3D local boxes with $L_x = L_y = 200 \rm ~ kpc$, $L_z = 250 \rm ~ kpc $ and a resolution of $256 \times 256 \times 320$ cells. The aim of their study was to reproduce local ICM conditions in various regions for both relaxed and disturbed galaxy clusters. For this reason, they performed four simulation runs with increasing levels of stratification corresponding to different Froude numbers ($Fr = \sqrt{1/ Ri}$). They used the Froude number to quantify the role of the buoyancy in each simulation. The initial velocity field was generated with the purpose to mimic a 3D Kolmogorov turbulence, with an injection scale $L = 43.6$ kpc. They also removed the compressive modes to generate merely solenoidal turbulence. 
   Given the initial conditions, the evolution of the velocity distribution is fully dictated by the Mach and Froude number. In this sense, these simulations are quite similar to those of \citeauthor{2020MNRAS.493.5838M}. 
  
  For comparison, we analysed the most stratified  cluster data cube at 2 Gyr and estimated all the relevant quantities, such as the Richardson number and the logarithmic density fluctuations. In table~\ref{tab:Flash_parameter} we compare our results with the parameters of the snapshot (S\&Z method), both analysing the whole cube (full data cube) and boxes with $80^3$ cells located in various spot of the cube itself. To estimate the Froude number we used the injection scale of their simulations, which is $L = 43.6$ kpc.
  
  \begin{center} 
    \begin{table}
    \begin{center}
    \begin{tabular}{|c|c|c|} 
    
    \hline
    & $N_{\rm BV} ({\rm Gyr}^{-1})$  &  Froude number  \\ \hline 
    
    S\&Z method & 10.6                 & 0.34  \\ \hline 
    
    Full data cube & $13.44 \pm 6.86 $      & $0.37 \pm 0.23$ \\ \hline
    
    1st box    & $9.08 \pm 2.98$        & $0.41 \pm 0.11$\\ \hline
    
    2nd box    & $9.21 \pm 3.04$        & $0.29 \pm 0.06$ \\ \hline
    
    3rd box    & $9.13 \pm 3.92$        & $0.45 \pm 0.13$ \\ \hline
    
    4th box    & $8.70 \pm 3.48$        & $0.39 \pm 0.09$ \\ \hline
    
    5th box    & $8.70 \pm 2.35$        & $0.36 \pm 0.06$ \\ \hline
    
    6th box    & $7.22 \pm 2.97$        & $0.47 \pm 0.21$ \\ \hline
    
    7th box    & $7.08 \pm 2.85$        & $0.44 \pm 0.14$ \\ \hline
    
    8th box    & $7.18 \pm 3.13$        & $0.45 \pm 0.17$ \\ \hline
    
    9th box    & $7.28 \pm 3.57$        & $0.48 \pm 0.17$ \\ \hline

    \end{tabular}
    \end{center} 
    \caption{\label{tab:Flash_parameter} Comparison between the results of our analysis, applied to the \cite{2019MNRAS.487.1072S} simulation, and their output parameters.}
    \end{table}
    \end{center}
    
\begin{figure}
        \centering
        \includegraphics[width= 0.5\textwidth]{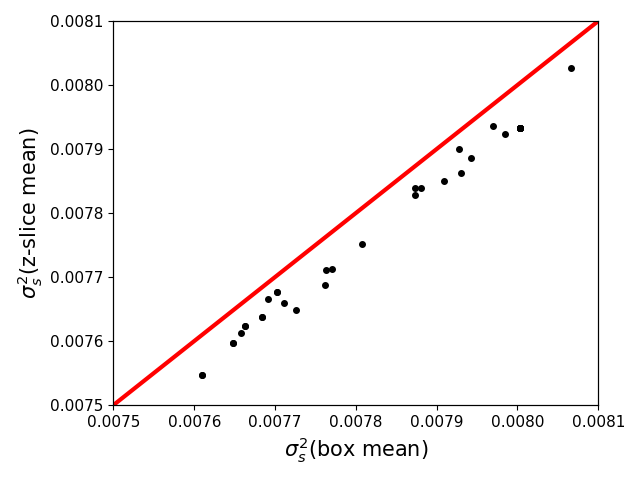}
        \caption {A comparison between the logarithmic density fluctuations estimated with a fixed scale filtering method and calculating the density mean in slices at a given $z$ coordinate. The red line is the bisector, which highlights the one-to-one relation.}
        \label{fig:d_test}
   \end{figure}

   After filtering for density fluctuations, we obtained the logarithmic perturbations both performing a fixed-scale filtering method, with a scale $L \approx 43.6$ kpc and computing the mean density on a slice at a given $z$ coordinate, which is the method that \citealt{2020MNRAS.493.5838M}, \citeyear{2021MNRAS.500.5072M} used in their studies. We plot the results in Fig.~\ref{fig:d_test} for a comparison between the two filtering algorithms. The black points are the results of the analysis performed in a box with $80^3$ cells. Finally, we estimated the mean logarithmic density fluctuation considering the full data cube, obtaining $\sigma^2_s = 0.048$. This result is in close
   agreement with Fig.~8 of \citet{2021MNRAS.500.5072M}, which predicts $\sigma^2_s \approx 0.04 - 0.05 $ for a Froude number $Fr \approx 0.3 - 0.4$. 
   In summary, our results  match reasonably well those in the literature, which instills confidence in our filtering methods, both for idealised as well as cosmological simulations.
   
     \section{Results}
     \label{sec:results}
     
     First, we focus on the relation between the density and velocity fluctuations in our cluster sample. Then, we investigate how the stratification can affect this relation and test the dependence of the density fluctuation on the Richardson number. A large part of the analysis was performed within the virial radius of the galaxy clusters in order to compare to observations.
     
     \subsection{The $\sigma_{\rho} - \sigma_{v} $ relation}
     
In observations, turbulence in galaxy clusters has been investigated using the relation between the gas density and velocity fluctuations. This has been followed up by numerical work in, both, cosmological and idealised simulations (\citealt{2014ApJ...788L..13Z}, \citealt{2014A&A...569A..67G}). The goal of those studies was to derive the amplitude of gas turbulent motions solely from the analysis of surface brightness fluctuations. Those studies found a one-to-one scaling relation suggesting that such a relation is valid in, both, perturbed and relaxed clusters. 
     
\begin{figure}
        \centering
        \includegraphics[width= 0.5\textwidth]{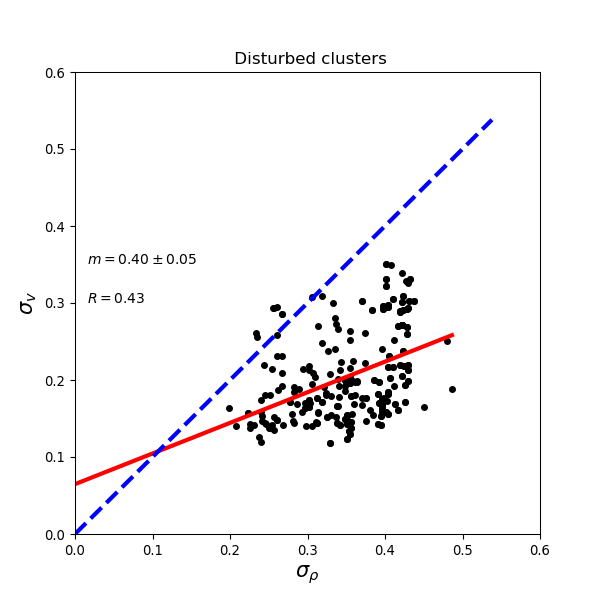}\quad\includegraphics[width= 0.5\textwidth]{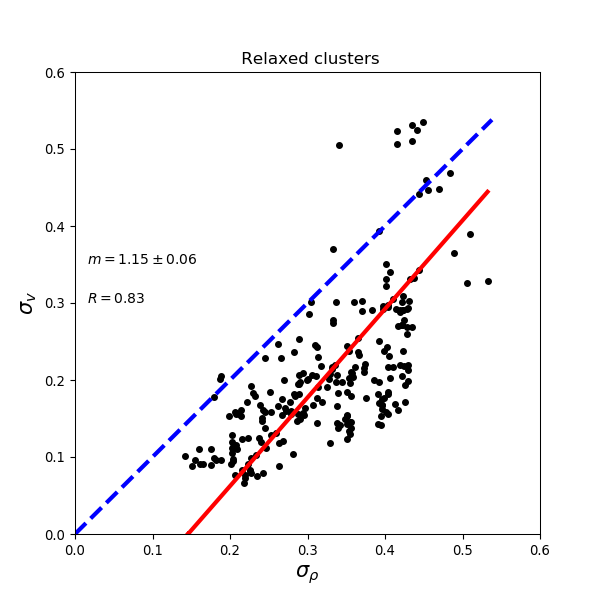}
        \caption{$\sigma_{\rho} - \sigma_v$ relation in our sample, which shows a dependence on the dynamical state of the galaxy cluster. The classification is based on the centroid shift parameter. The threshold value is $w_c = 0.0067$. The blue line denotes the 1-to-1 relation found by \citet{2014ApJ...788L..13Z}.}
        \label{fig:v_d_relation}
     \end{figure}

     Fig.~\ref{fig:v_d_relation} shows the $\delta \rho/\rho - \delta v / c_s$ relation in our sample for, both, relaxed and disturbed clusters. The classification of the dynamical state is based on the centroid shift parameter, $w$, which measures the offset between the centre of mass and the position of the gas density peak of the cluster (\citealt{2013AstRv...8a..40R}). The centroid shift of each cluster, $\langle w_i \rangle$, was estimated averaging the parameter values of multiple lines of sight. Galaxy clusters are then classified comparing $\langle w_i \rangle$ with  the median of the centroid shift of the sample, which is $w_c=0.0067$ (which is half of the value used in \citealt{2011JApA...32..519C}), with relaxed clusters having $w \leq w_c$. The relation was obtained within the virial volume by applying a fixed scale filtering method with scale $L$ = 300 kpc and averaging the density and velocity fluctuations over 600 kpc wide boxes (see Sec.~\ref{sec:filtering}). The outcome is a linear relation that depends on the dynamical state. Relaxed clusters have a slope $m = 1.15 \pm 0.06$, where $\sigma_v \propto m\sigma_{\rho}$, consistent with \citet{2014ApJ...788L..13Z}. Contrary to expectations, disturbed clusters show a flatter relation, with $m = 0.40 \pm 0.05$.
     
     In order to test the goodness of the fit, we used the Pearson's correlation coefficient
     
     \begin{equation}
       R = \frac{\sum \limits_{i} (x_i - \bar{x}) (y_i - \bar{y})} {\sqrt{ \sum \limits_{i} (x_i - \bar{x})^2 \sum \limits_{i} (y_i - \bar{y})^2 }},
   \end{equation}
   where $\bar{x}$ and $\bar{y}$ are the means of the variables used for the fit. $R$ is a measure of linear correlation between two sets of data, but it ignores other types of relationship. The correlation coefficient ranges between -1 and 1. These limits represent a perfect (anti)correlation, while a value of 0 implies that the variables are uncorrelated. For this fit, we obtained $R = 0.43 $ and $R = 0.82$ for perturbed and relaxed clusters, respectively.
   
     \begin{figure}
        \centering
        \includegraphics[width= 0.5\textwidth]{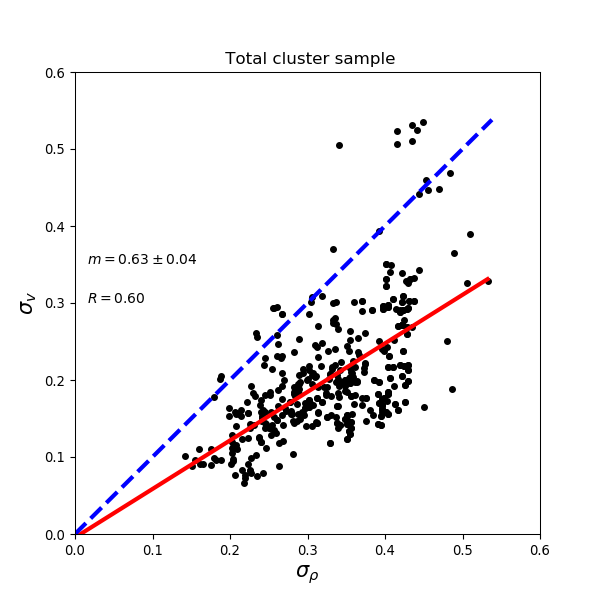}
        \caption{$\sigma_{\rho} - \sigma_v$ relation for the total cluster sample.The blue line denotes the 1-to-1 relation found by \citet{2014ApJ...788L..13Z}.}
        
        \label{fig:total}
        
    \end{figure}
     
\begin{figure}[]
   	\centering
   	\includegraphics[width= 0.5\textwidth]{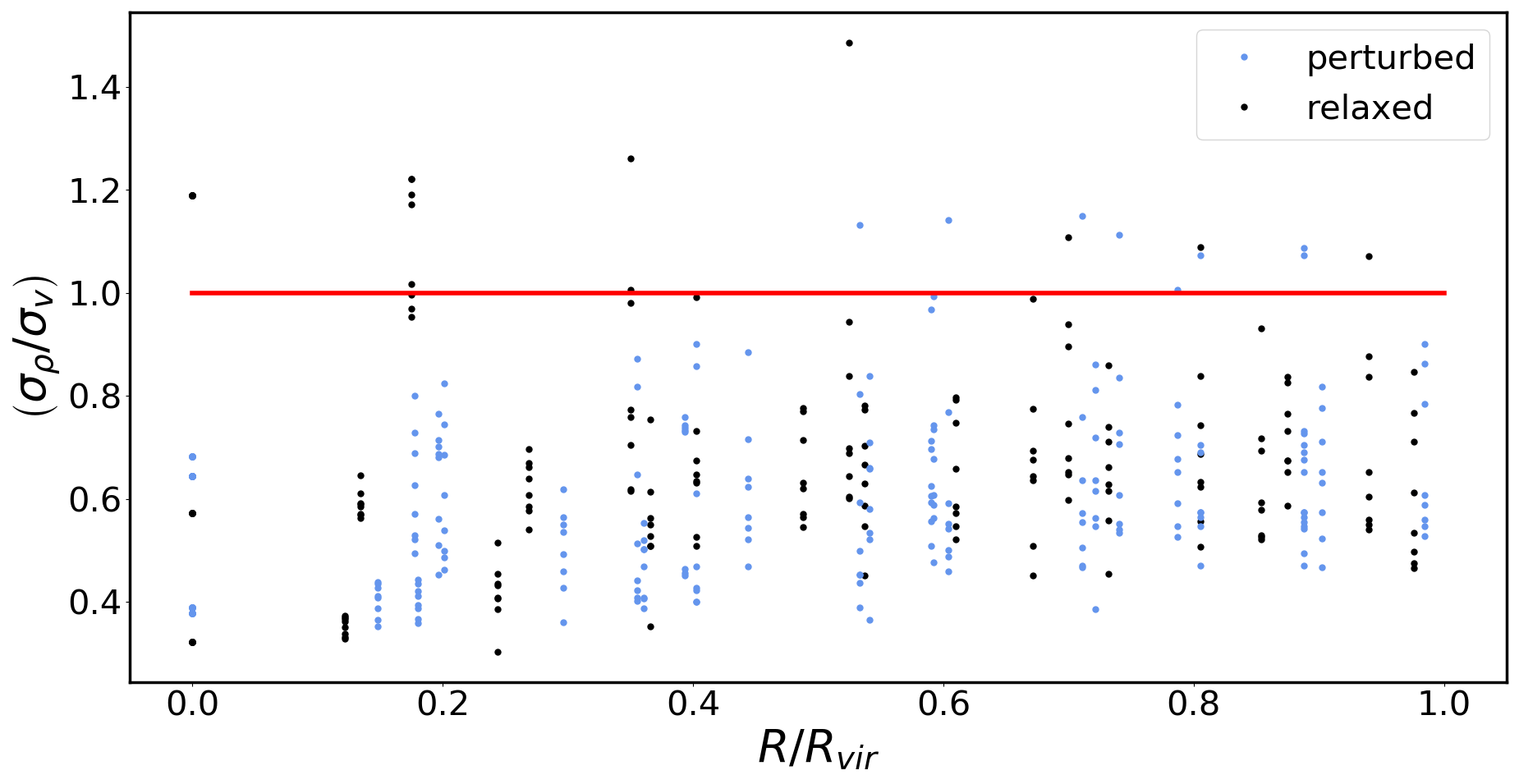}\quad\includegraphics[width= 0.5\textwidth]{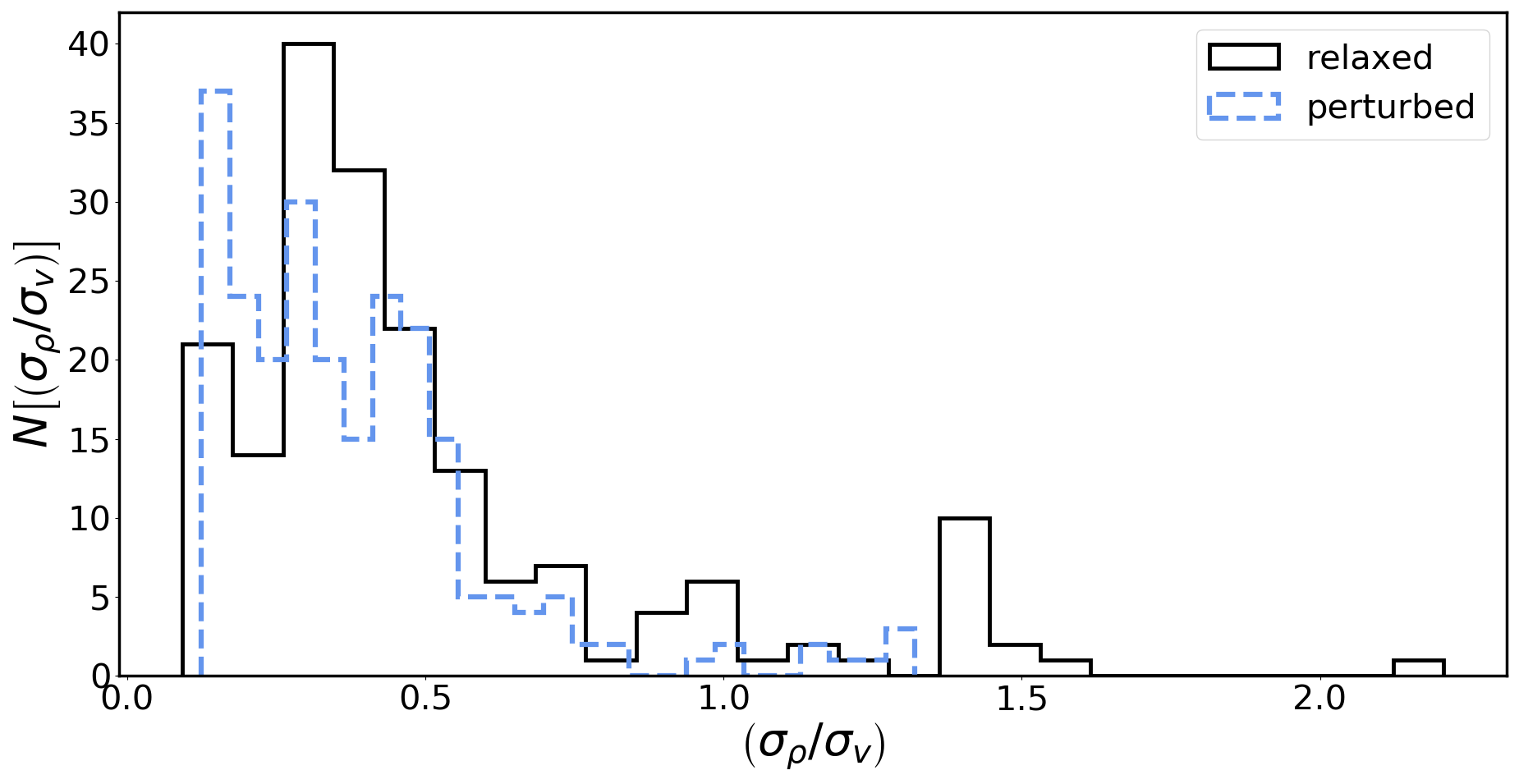}
   	\caption{Top panel: ratio of the amplitudes of density and velocity fluctuations as function of the distance from the cluster centre. 
   	   Bottom panel: the distribution of the ratio between density and velocity rms. The black points (line) correspond to relaxed clusters and blue points (line) to the perturbed ones. The classification into relaxed and perturbed clusters is based on the centroid shift parameter, whose discriminating value is $w_c = 0.0067$.
   	   }
   	\label{fig:ratio_radial_profile}
   \end{figure}

\begin{figure}
        \centering
        \includegraphics[width=0.47\textwidth]{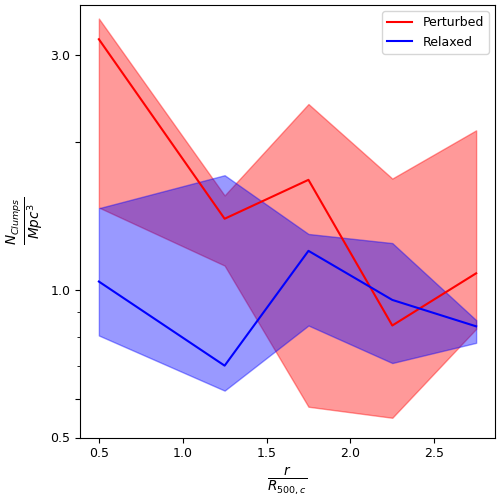}
        \caption{Radial profile of the number density of clumps for both relaxed (blue) and perturbed (red) galaxy clusters. The shadowed region represent the 16th and 84th percentiles of the bin’s number density distribution.}
        \label{fig:clump_profile}
  \end{figure}

     Moreover, in Fig.~\ref{fig:ratio_radial_profile} we show the distribution of the ratio of the rms values of density and velocity fluctuations which depends depends slightly on the cluster-centric radius, regardless of the dynamical state of the cluster. Blue and black lines (or points) are related to relaxed and disturbed clusters, respectively, while the red line represents the relation found by \citet{2014ApJ...788L..13Z}.\newline
     The Itasca simulation are non-radiative. This implies that some features of cool-core clusters, such as the peak (drop) in the density (temperature) radial profile in the centre of galaxy cluster, are missing. For this reason, it is not easy to make a proper classification, dividing our clusters in relaxed or disturbed, in this case. Due to this uncertainty, we performed the fit of the $\sigma_{\rho} - \sigma_v$ relation also considering the total cluster sample (Fig.~\ref{fig:total}). We found a flatter relation, compared to a one-to-one relation, with a slope $m = 0.63 \pm 0.04$ and a Pearson's correlation coefficient $R = 0.60$.\newline
     Finally, we noticed that some fluctuations that are not driven by turbulence might affect the $\sigma_{\rho} - \sigma_v$ relation, even after filtering. As a result, some of the $\sigma_v$ and $\sigma_{\rho}$ values in Fig.~\ref{fig:v_d_relation} might be caused by such contamination. \citet{2021arXiv210201096A} recently studied the radial profile of the clumping factor and the number density of clump in the Itasca sample. Here, we performed a similar analysis. We divided the inner region (within $3R_{500}$) into six bins. For each of them we calculated the number of clumps as a function of the respective volume. The mass threshold for the clumps is $10^{8} M_{\odot}$ and we set their maximum size to be 300 kpc. Fig.~\ref{fig:clump_profile} shows the radial profile of the median clump number density. The shadowed regions represents the 16th and 84th clump number density distribution percentile boundaries. The number of clumps in the innermost region is higher in the perturbed clusters, making it less straightforward to study the relation between density and velocity fluctuations. However, the presence of self-gravitating structures is a problem for X-ray observations, too. In observations the identification of clumps is even more complicated, due to their low surface brightness.
     A solution might be to perform the analysis using a smaller filtering scale, in order to filter out also the smallest clumps. Nevertheless, the driver of turbulence in this simulation is the radial accretion. Thus, a smaller filtering scale would not be physically motivated since the injection scales generated by such event are likely larger than 200 kpc \citep{2018MNRAS.481L.120V,2020MNRAS.495..864A}.

   \subsection{The $Ri - \sigma^2_s$ correlation}
   \label{sec:Ri_sigma}
   
\begin{figure}[h!]
        \centering
        \includegraphics[width= 0.50\textwidth]{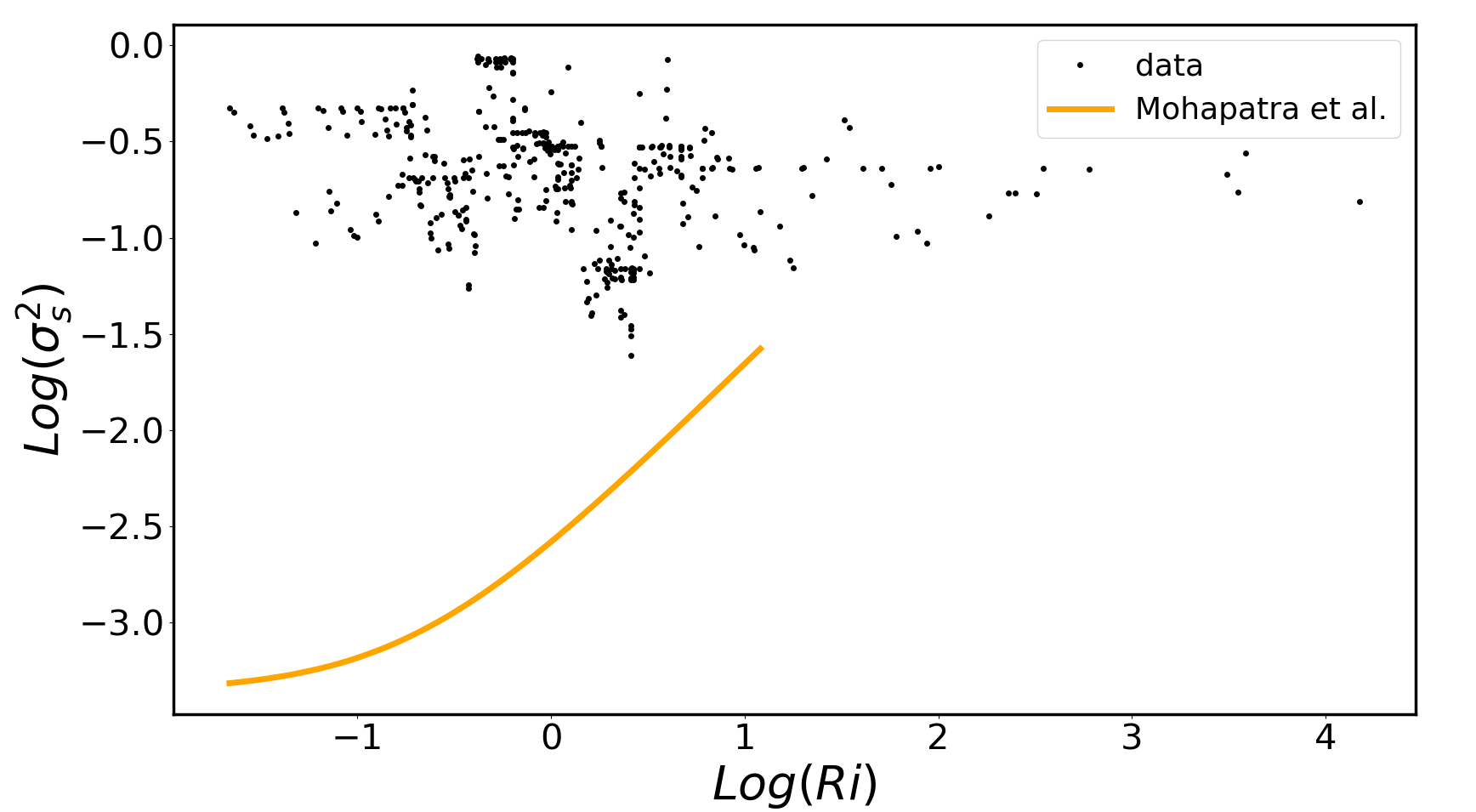}
        \caption{$Ri$ - $\sigma^2_s$ plot. Density perturbations are independent of the Richardson number and much higher than those found by \citeauthor{2020MNRAS.493.5838M} (orange line).
        The analysis was performed on the total cluster sample.}
        \label{fig:Ri_sigma}
  \end{figure}

   In the previous section we discussed the relation between density and velocity perturbations (Fig.~\ref{fig:v_d_relation}). However, the stratification and, consequently, the buoyancy make the interpretation of gas fluctuations less straightforward. Hence, we computed the mean of the Richardson number and the logarithmic density fluctuations over 600 kpc wide boxes using a filtering method with a fixed scale of $L = 300 \rm~kpc $ (see Sec.~\ref{sec:filtering}). In \citet{2020MNRAS.493.5838M}, the authors found a tight relation between density fluctuations and the Richardson number (highlighted by the orange line in Fig.~\ref{fig:Ri_sigma}), which we cannot detect in our sample (black points in Fig.~\ref{fig:Ri_sigma}). In particular, our simulated clusters show higher density fluctuations, up to three orders of magnitude, for $Ri < 1$, and a lower limit on $\log_{10}(\sigma^2_s) \approx -1.5$, regardless of $Ri$.
   
\begin{figure}
        \centering
        \includegraphics[width= 0.5\textwidth]{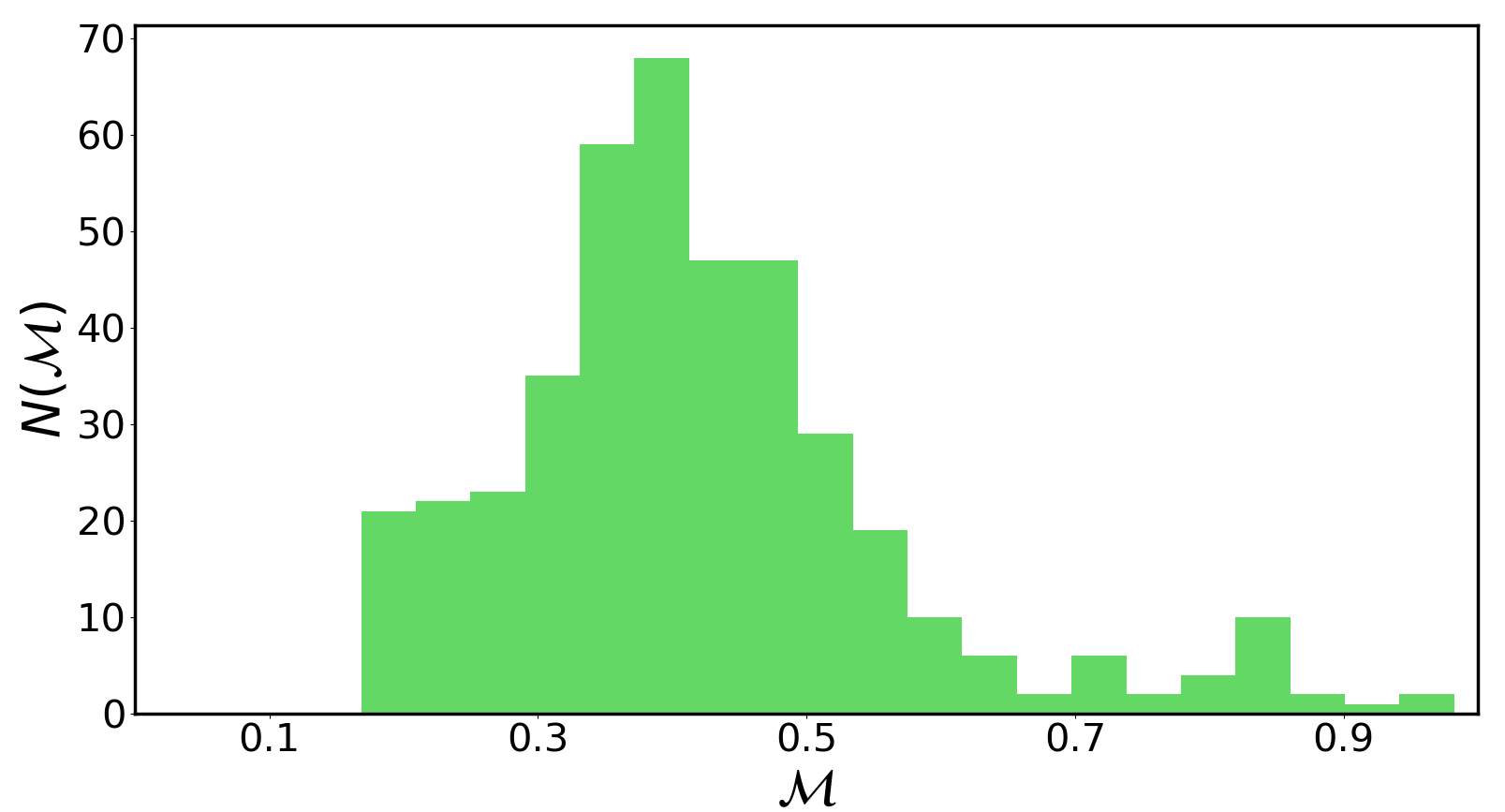}
        \caption {Distribution of the Mach numbers in our cluster sample, using averaging over 600 kpc. The tail of the distribution at high Mach numbers is caused by local, supersonic turbulence.}
        \label{fig:Mach_number}
  \end{figure}

We have already presented in Sec.~\ref{fig:power_spectra} our tests on the reliability of the density fluctuations computed in the Itasca cluster. Even though our distribution does not perfectly match that of \citet{2017ApJ...843L..29E}, the deviation in the surface brightness fluctuations is not sufficient to obtain density fluctuations similar to those of \citet{2020MNRAS.493.5838M}. Moreover, we have also shown in Sec.~\ref{sec:test} that when applied to a similar (idealised) simulation setup, our pipeline correctly recovers the expected density fluctuations and Froude or Richardson number.
We are therefore led to the conclusion that the setup in \citealt{2020MNRAS.493.5838M} and \citeyear{2021MNRAS.500.5072M} is much more idealised than more realistic ICM as produced by the Itasca sample, both concerning the solenoidal forcing of turbulence and the absence of self-gravitating gas substructures and bulk motions associated with accretion.
In our cluster sample, the accretion of clumps is ubiquitous and it produces both solenoidal and compressive turbulence. This results in a distribution of flow velocities with typical averaged (on 600 kpc wide boxes) Mach numbers of $\approx 0.4-0.5$ and likely locally supersonic motions (see Fig.~\ref{fig:Mach_number}). Another crucial difference is the distribution of the logarithmic density fluctuations, which is found to be significantly different from a log-normal distribution in our work. 
 In summary, in our simulations we find that density fluctuations do not depend on $Ri$. Therefore, estimating the density perturbations as the variance of the probability distribution function, and relating it to the physical properties of the underlying turbulent flow, appears to be challenging in realistic situations, even without taking into account observational effects and sampling issues.

  \subsection{The $Ri - \beta$ correlation}
  
\begin{figure}
        \centering
        \includegraphics[width= 0.5\textwidth]{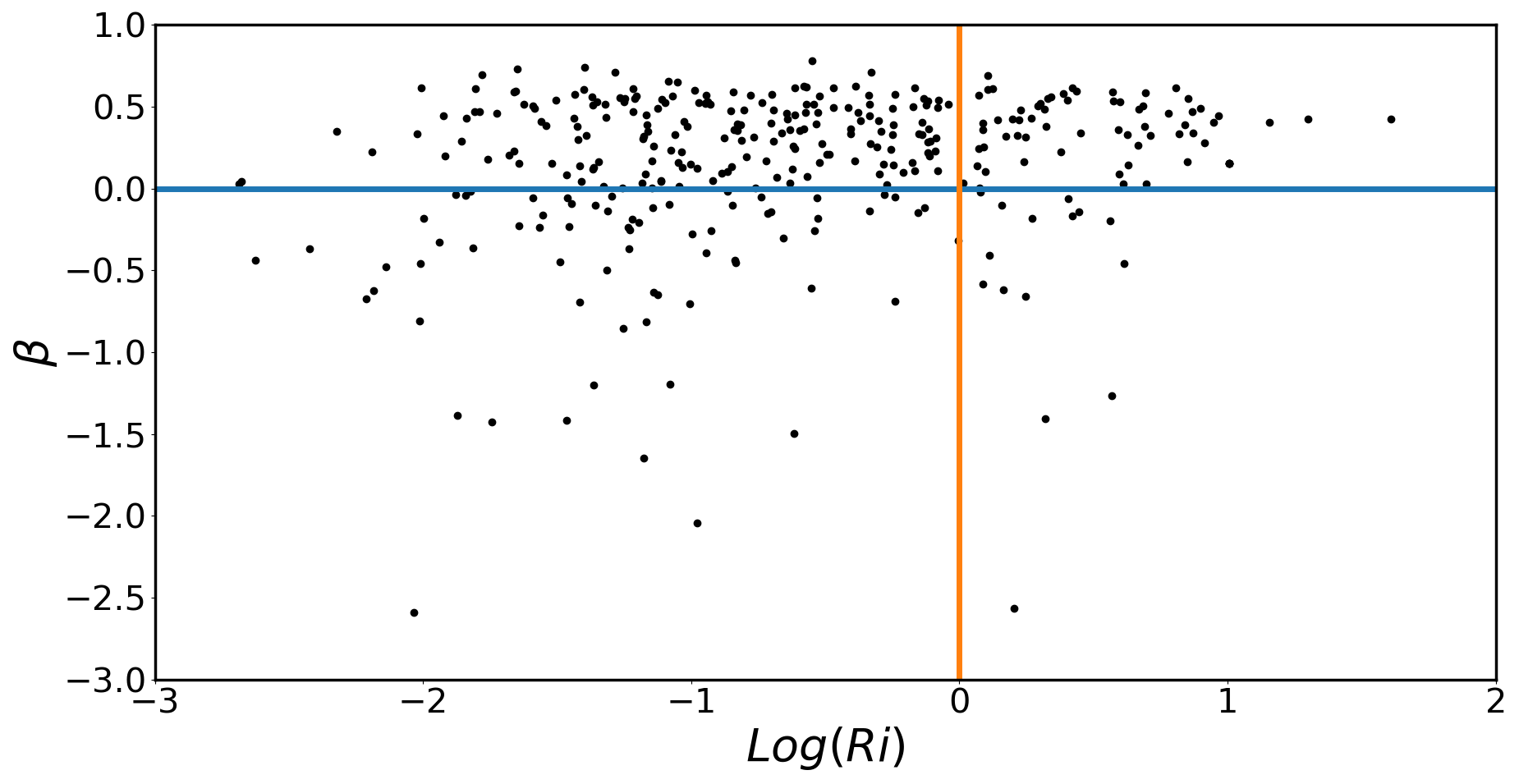}
        \caption {$Ri - \beta$ relation in our cluster sample. The radial bias is an imprint that the dynamics is dominated by radial motions, driven by the accretion. In this sense, the Richardson number does not have a predictive power about the real dynamics of the ICM. The orange line is used to highlight the turbulence (left) and buoyancy (right) dominated regime, while the blue line distinguishes a radially ($\beta > 0$) from a tangentially ($\beta < 0$) bias.}
        \label{fig:Ri_beta}
  \end{figure}

  The stratification and buoyancy might also affect the velocity distribution of the plasma, suppressing the radial motion. As a result, in the buoyancy-dominated regime ($Ri > 1$), the gas dynamics should be dominated by tangential motions (see \citealt{2007JFM...585..343B}, \citealt{2020MNRAS.493.5838M}, \citeyear{2021MNRAS.500.5072M}). Therefore, we look for a relation between the Richardson number and the anisotropy parameter. Since for $Ri < 1$ the hypotheses of homogeneity and isotropy are still valid, we should expect a mean $\beta \sim 0$. On the other hand, a mean $\beta < 0$ should arise when we move to buoyancy-dominated regimes. However, looking at Fig.~\ref{fig:Ri_beta}, we see that the dynamics of the ICM is biased by radial motions despite the stratification. As already noticed in previous work on this sample \citep[e.g.][]{2018MNRAS.481L.120V}, this ubiquitous dynamical feature of the ICM is most likely a consequence of radial accretion flows that are important drivers of ICM turbulence.
  Finally, we have explored the sensitivity of our results on the choice of the filtering scale and box size, by considering larger and smaller filtering scales. We find that qualitatively the results still hold (see Appendix 1 for more details).
  
  The important conclusion is that the Richardson number is not able to characterise the effect of ICM stratification for cosmologically evolving clusters.
  Since accretion occurs largely via the infall of gas clumps whose motion is predominantly radial  \citep[e.g.][]{2021arXiv210201096A}, a predominantly radial (turbulent) velocity is found at most cluster-centric distances in all clusters of our sample. This radial turbulence is unlike the stratified turbulence expected for a $Ri \geq 1$ flow.
  
  \subsection{Radial profile of the power spectra slope}
  
  The velocity power spectrum quantifies the strength of turbulent fluctuations on different scales, and the slope of the power spectrum is customarily compared to Kolmogorov's theory.
  
\begin{figure}
        \centering
        \includegraphics[width= 0.5\textwidth]{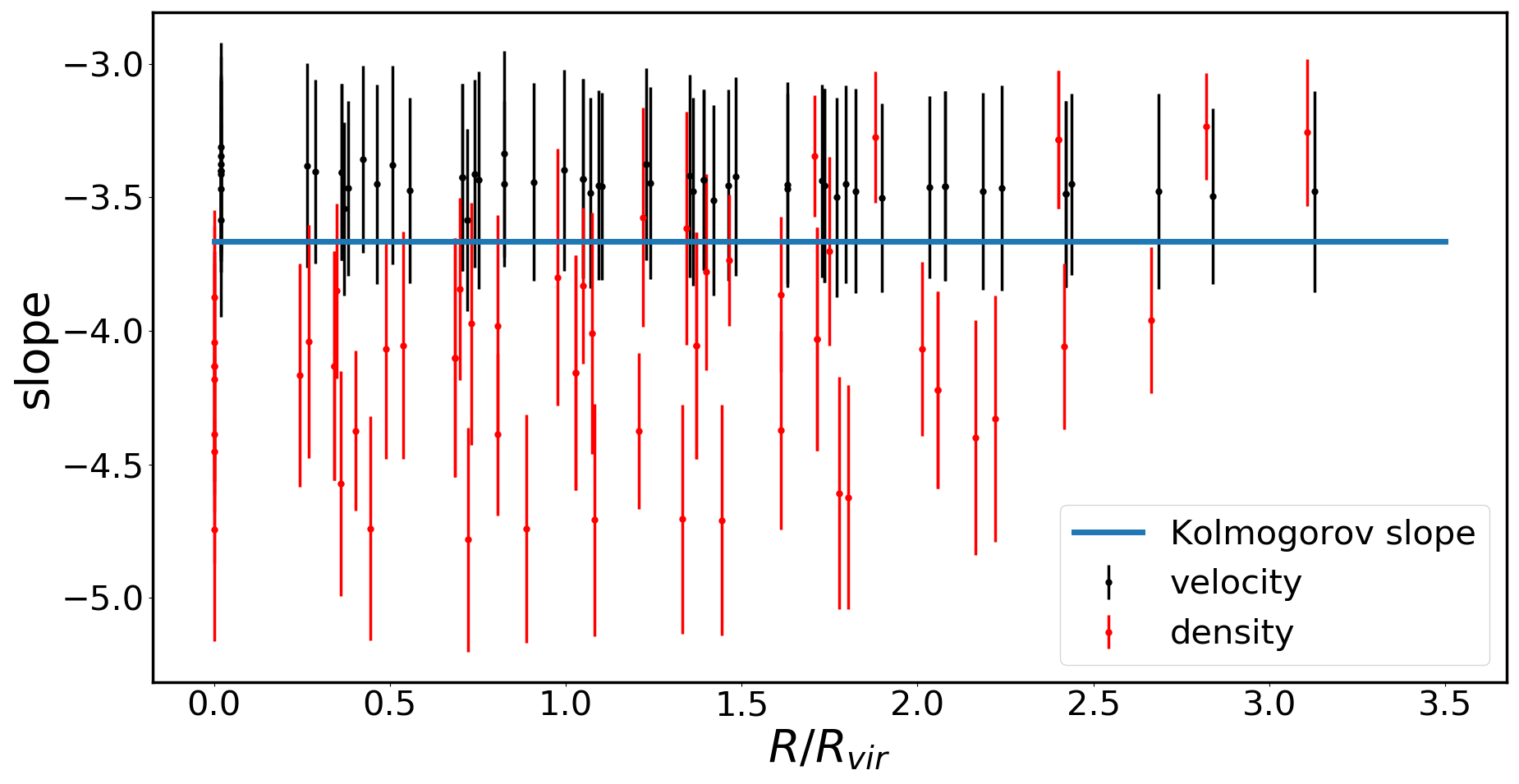}
        \caption{Radial profile of the slope of density and velocity power spectra in our cluster sample. The velocity slope is in agreement with Kolmogorov's theory and shows a constant profile as well as the slope of the density spectra, which is steeper. The blue line traces Kolmogorov's prediction for the velocity slope of a 3D spectrum.}
        \label{fig:slope_radial_profile}
  \end{figure}

  We compute the radial profile of slope of the density and velocity power spectra in the Itasca cluster sample, which is shown in Fig.~\ref{fig:slope_radial_profile}. The error-bars represent the $1\sigma$ error, which were estimated as the standard error of the weighted mean $\sigma = \left( \sqrt{\sum \limits_{i} \frac{1}{\sigma_i^2}} \right)^{-1}$, where $\sigma_i$ are the $1\sigma$ errors of each power spectrum at a given distance from the centre of the cluster. The slope of the velocity spectrum shows a constant radial profile and agrees with the Kolmogorov's model, indicated by the blue line. On the other hand, the density power spectra have steeper slopes, albeit with a bigger scatter. The slope of the velocity spectrum is flatter compared to \citet{2017MNRAS.464..210V}, who used the same cluster sample. The reason for this difference is explained by the varying normalisation of the power spectrum, which decreases as we move to larger radii (see Fig.~\ref{fig:power_spectra}). While in previous work we measured a unique power spectrum across a large volume (hence mixing local spectra with a slightly decreasing normalisation going to the cluster outskirts), here we computed spectra in smaller boxes.  
  On scales of $\leq 600 \rm ~ kpc$ the turbulence appears even more Kolmogorov-like, despite the presence of radial biases and a non-stationary forcing.
  At the same time, the larger scatter in the slope of the density power spectrum at all radii questions whether any simple relation can be robustly derived between the two quantities. 
  
  
   \section{Discussion and conclusion}
   
   \label{sec:conclusion}
   
   We have investigated whether the turbulent gas velocity fluctuations in the ICM can be constrained using projected X-ray surface brightness fluctuations. 
   Previous work \citep{2014A&A...569A..67G, 2014ApJ...788L..13Z, 2004A&A...426..387S, 2012MNRAS.421..726S} have already shown that 
   the pattern of surface brightness fluctuations can, under certain limiting assumptions, be translated into density fluctuations.
   However, in a stratified ICM, the $\delta \rho - \delta v$ relation is affected by buoyancy effects. \citet{2020MNRAS.493.5838M}, (\citeyear{2021MNRAS.500.5072M}) further derived an accurate relation between the two, parameterised by the Richardson number and the Mach number of the flow, in the very simplified case of turbulence-in-a-box simulations. 
   So far, such analyses have been limited to idealised simulations or to the innermost cluster regions. Here, we used the Itasca cluster sample, which is a set of galaxy clusters simulated with the cosmological code ENZO with a uniform comoving spatial resolution of about $\approx 20$ kpc. In order to compare the results to observations, we limited our analysis of the ICM to within the virial radius.
   
   Fig.~\ref{fig:power_spectra} shows that while the density fluctuations increase as we move to larger radii, the velocity power spectra remain fairly constant. The increasing clumping factor in the outer region of galaxy clusters is the likely driver for this trend. 
   As a consequence, a more accurate algorithm to mask clumps is required in observational and numerical analyses. This will be particularly challenging for observations, where the masking small-scale and faint clumps will remain limited.
   We can summarise our results as follows:
   
   \renewcommand{\labelitemi}{$\bullet$}
   \begin{itemize}
   
       \item Consistent with previous work (\citealt{2014ApJ...788L..13Z}, \citealt{2014A&A...569A..67G}),
       we find that a linear relation exists between the standard deviation of the density contrast, $\sigma_{\rho}$ and the normalised velocity fluctuations, $\sigma_{v}$, albeit with a significant scatter across the sample.
       However, the slope of this correlation does not generally agree with the findings of the aforementioned studies. For the whole Itasca sample (Fig.~\ref{fig:total})
       it results $\sigma_{v} \approx  0.6\sigma_{\rho}$, with smaller velocity fluctuations than expected by applying the \citet{2014A&A...569A..67G} relation. 
       
       When we separate the sample in relaxed and perturbed clusters, we find that relaxed objects show a steeper slope, consistent with 1, and again a large scatter. Perturbed clusters follow instead a flatter, weaker relation, $\sigma_{v} \approx  0.4\sigma_{\rho}$, with a low correlation coefficient.
       
        
        Fig.~\ref{fig:clump_profile} shows that  this relation may be contaminated by unfiltered clumps. This demonstrates the problem in using this relation to infer the turbulence in realistic clusters since this could lead to incorrect estimates, for example, of the turbulent heating rate (\citealt{2014Natur.515...85Z}) and the kinetic energy available to explain radio halos ($P_H \propto \rho \sigma_v^3$, e.g. \citealt{2017ApJ...843L..29E}).
       
       \item Although buoyancy is the dominant process in the largest part of the cluster volume (Fig.~\ref{fig:Ri_profile}), we did not find a strong relation between the logarithmic density fluctuations and the Richardson number. The results in Sec.~\ref{sec:Ri_sigma} show that our fluctuations are much higher compared to those of \citet{2020MNRAS.493.5838M}, (\citeyear{2021MNRAS.500.5072M}), who performed an analogous analysis in more idealised simulations,  and are independent of the Richardson number. We attribute the difference to the unavoidable presence of compressive turbulence, consistent with previous work \citep[e.g.][]{2015ApJ...800...60M,2017MNRAS.464..210V, 2021MNRAS.504..510V} . Furthermore, the turbulent anisotropy parameter shows the effects of substantial radial accretion (see Fig.~\ref{fig:Ri_beta}), which appears to be the dominant driver for the gas fluctuations in our simulations \citep[e.g.][]{2021arXiv210201096A}.
       
       \item Turbulence in the ICM is usually assumed to be homogeneous and isotropic. We show that this may not be valid, even if the spectra are found to follow Kolmogorov's scaling. This is primarily caused by radial accretion that dominates the dynamics of the ICM (see Fig.~\ref{fig:Ri_beta}). In cluster cores, jets from central radio galaxies may produce a similar effect (\citealt{2015ASSL..407..599B}).
       
   \end{itemize}
 Finally, we point out that our cosmological simulations are optimised for a uniform resolution across the cluster volume and cannot attain the high spatial resolution of observations in cluster cores. Therefore, our simulations do not extend to scales below $\leq 10 \rm ~kpc$ probed by some X-ray observations of nearby clusters of galaxies. 

\section*{ACKNOWLEDGEMENTS}

The authors thank Xun Shi for providing data for subsection \ref{sec:test}. The cosmological simulations described in this work were performed using the {\enzo} code (http://enzo-project.org), which is the product of a collaborative effort of scientists at many universities and national laboratories. 
F.V. acknowledges financial support from the European Union's Horizon 2020 program under the ERC Starting Grant 'MAGCOW', no. 714196. Work by TWJ was supported by the US National Science Foundation grant AST1714205. MB acknowledges support by the Deutsche Forschungsgemeinschaft (DFG, German Research Foundation) under Germany's Excellence Strategy -- EXC 2121 ``Quantum Universe'' --  390833306, Germany’s Excellence Strategy -- EXC-2094 ``Origins'' -- 390783311.

\bibliographystyle{aa}
\bibliography{marco}

\begin{thebibliography}{}
\makeatletter
\relax
\def\mn@urlcharsother{\let\do\@makeother \do\$\do\&\do\#\do\^\do\_\do\%\do\~}
\def\mn@doi{\begingroup\mn@urlcharsother \@ifnextchar [ {\mn@doi@}
  {\mn@doi@[]}}
\def\mn@doi@[#1]#2{\def\@tempa{#1}\ifx\@tempa\@empty \href
  {http://dx.doi.org/#2} {doi:#2}\else \href {http://dx.doi.org/#2} {#1}\fi
  \endgroup}
\def\mn@eprint#1#2{\mn@eprint@#1:#2::\@nil}
\def\mn@eprint@arXiv#1{\href {http://arxiv.org/abs/#1} {{\tt arXiv:#1}}}
\def\mn@eprint@dblp#1{\href {http://dblp.uni-trier.de/rec/bibtex/#1.xml}
  {dblp:#1}}
\def\mn@eprint@#1:#2:#3:#4\@nil{\def\@tempa {#1}\def\@tempb {#2}\def\@tempc
  {#3}\ifx \@tempc \@empty \let \@tempc \@tempb \let \@tempb \@tempa \fi \ifx
  \@tempb \@empty \def\@tempb {arXiv}\fi \@ifundefined
  {mn@eprint@\@tempb}{\@tempb:\@tempc}{\expandafter \expandafter \csname
  mn@eprint@\@tempb\endcsname \expandafter{\@tempc}}}

\bibitem[\protect\citeauthoryear{{Angelinelli}, {Vazza}, {Giocoli}, {Ettori},
  {Jones}, {Brunetti}, {Br{\"u}ggen}  \& {Eckert}}{{Angelinelli}
  et~al.}{2020}]{2020MNRAS.495..864A}
{Angelinelli} M.,  {Vazza} F.,  {Giocoli} C.,  {Ettori} S.,  {Jones} T.~W.,
  {Brunetti} G.,  {Br{\"u}ggen} M.,   {Eckert} D.,  2020, \mn@doi [\mnras]
  {10.1093/mnras/staa975}, \href
  {https://ui.adsabs.harvard.edu/abs/2020MNRAS.495..864A} {495, 864}

\bibitem[\protect\citeauthoryear{{Angelinelli}, {Ettori}, {Vazza}  \&
  {Jones}}{{Angelinelli} et~al.}{2021}]{2021arXiv210201096A}
{Angelinelli} M.,  {Ettori} S.,  {Vazza} F.,   {Jones} T.~W.,  2021, arXiv
  e-prints, \href {https://ui.adsabs.harvard.edu/abs/2021arXiv210201096A} {p.
  arXiv:2102.01096}

\bibitem[\protect\citeauthoryear{{Ar{\'e}valo}, {Churazov}, {Zhuravleva},
  {Forman}  \& {Jones}}{{Ar{\'e}valo} et~al.}{2016}]{2016ApJ...818...14A}
{Ar{\'e}valo} P.,  {Churazov} E.,  {Zhuravleva} I.,  {Forman} W.~R.,   {Jones}
  C.,  2016, \mn@doi [\apj] {10.3847/0004-637X/818/1/14}, \href
  {https://ui.adsabs.harvard.edu/abs/2016ApJ...818...14A} {818, 14}

\bibitem[\protect\citeauthoryear{{Bonafede} et~al.,}{{Bonafede}
  et~al.}{2018}]{2018MNRAS.478.2927B}
{Bonafede} A.,  et~al., 2018, \mn@doi [\mnras] {10.1093/mnras/sty1121}, \href
  {https://ui.adsabs.harvard.edu/abs/2018MNRAS.478.2927B} {478, 2927}

\bibitem[\protect\citeauthoryear{{Brethouwer}, {Billant}, {Lindborg}  \&
  {Chomaz}}{{Brethouwer} et~al.}{2007}]{2007JFM...585..343B}
{Brethouwer} G.,  {Billant} P.,  {Lindborg} E.,   {Chomaz} J.~M.,  2007,
  \mn@doi [Journal of Fluid Mechanics] {10.1017/S0022112007006854}, \href
  {https://ui.adsabs.harvard.edu/abs/2007JFM...585..343B} {585, 343}

\bibitem[\protect\citeauthoryear{{Br{\"u}ggen} \& {Kaiser}}{{Br{\"u}ggen} \&
  {Kaiser}}{2002}]{2002Natur.418..301B}
{Br{\"u}ggen} M.,  {Kaiser} C.~R.,  2002, \mn@doi [\nat] {10.1038/nature00857},
  \href {https://ui.adsabs.harvard.edu/abs/2002Natur.418..301B} {418, 301}

\bibitem[\protect\citeauthoryear{{Br{\"u}ggen} \& {Vazza}}{{Br{\"u}ggen} \&
  {Vazza}}{2015}]{2015ASSL..407..599B}
{Br{\"u}ggen} M.,  {Vazza} F.,  2015, {Turbulence in the Intracluster Medium}.
p.~599, \mn@doi{10.1007/978-3-662-44625-6_21}

\bibitem[\protect\citeauthoryear{{Brunetti} \& {Jones}}{{Brunetti} \&
  {Jones}}{2014}]{bj14}
{Brunetti} G.,  {Jones} T.~W.,  2014, \mn@doi [International Journal of Modern
  Physics D] {10.1142/S0218271814300079}, \href
  {http://adsabs.harvard.edu/abs/2014IJMPD..2330007B} {23, 1430007}

\bibitem[\protect\citeauthoryear{{Bryan} et~al.,}{{Bryan}
  et~al.}{2014}]{2014ApJS..211...19B}
{Bryan} G.~L.,  et~al., 2014, \mn@doi [\apjs] {10.1088/0067-0049/211/2/19},
  \href {https://ui.adsabs.harvard.edu/abs/2014ApJS..211...19B} {211, 19}

\bibitem[\protect\citeauthoryear{{Cassano}, {Brunetti}  \& {Venturi}}{{Cassano}
  et~al.}{2011}]{2011JApA...32..519C}
{Cassano} R.,  {Brunetti} G.,   {Venturi} T.,  2011, \mn@doi [Journal of
  Astrophysics and Astronomy] {10.1007/s12036-011-9117-1}, \href
  {https://ui.adsabs.harvard.edu/abs/2011JApA...32..519C} {32, 519}

\bibitem[\protect\citeauthoryear{{Churazov} et~al.,}{{Churazov}
  et~al.}{2012}]{2012MNRAS.421.1123C}
{Churazov} E.,  et~al., 2012, \mn@doi [\mnras]
  {10.1111/j.1365-2966.2011.20372.x}, \href
  {https://ui.adsabs.harvard.edu/abs/2012MNRAS.421.1123C} {421, 1123}

\bibitem[\protect\citeauthoryear{{Colella} \& {Woodward}}{{Colella} \&
  {Woodward}}{1984}]{1984JCoPh..54..174C}
{Colella} P.,  {Woodward} P.~R.,  1984, \mn@doi [Journal of Computational
  Physics] {10.1016/0021-9991(84)90143-8}, \href
  {https://ui.adsabs.harvard.edu/abs/1984JCoPh..54..174C} {54, 174}

\bibitem[\protect\citeauthoryear{{De Young}}{{De
  Young}}{2010}]{2010ApJ...710..743D}
{De Young} D.~S.,  2010, \mn@doi [\apj] {10.1088/0004-637X/710/1/743}, \href
  {https://ui.adsabs.harvard.edu/abs/2010ApJ...710..743D} {710, 743}

\bibitem[\protect\citeauthoryear{{Dolag}, {Vazza}, {Brunetti}  \&
  {Tormen}}{{Dolag} et~al.}{2005}]{2005MNRAS.364..753D}
{Dolag} K.,  {Vazza} F.,  {Brunetti} G.,   {Tormen} G.,  2005, \mn@doi [\mnras]
  {10.1111/j.1365-2966.2005.09630.x}, \href
  {https://ui.adsabs.harvard.edu/abs/2005MNRAS.364..753D} {364, 753}

\bibitem[\protect\citeauthoryear{{Eckert}, {Gaspari}, {Vazza}, {Gastaldello},
  {Tramacere}, {Zimmer}, {Ettori}  \& {Paltani}}{{Eckert}
  et~al.}{2017}]{2017ApJ...843L..29E}
{Eckert} D.,  {Gaspari} M.,  {Vazza} F.,  {Gastaldello} F.,  {Tramacere} A.,
  {Zimmer} S.,  {Ettori} S.,   {Paltani} S.,  2017, \mn@doi [\apjl]
  {10.3847/2041-8213/aa7c1a}, \href
  {https://ui.adsabs.harvard.edu/abs/2017ApJ...843L..29E} {843, L29}

\bibitem[\protect\citeauthoryear{{Eckert} et~al.,}{{Eckert}
  et~al.}{2019}]{2019A&A...621A..40E}
{Eckert} D.,  et~al., 2019, \mn@doi [\aap] {10.1051/0004-6361/201833324}, \href
  {https://ui.adsabs.harvard.edu/abs/2019A&A...621A..40E} {621, A40}

\bibitem[\protect\citeauthoryear{{Ettori} et~al.,}{{Ettori}
  et~al.}{2019}]{2019A&A...621A..39E}
{Ettori} S.,  et~al., 2019, \mn@doi [\aap] {10.1051/0004-6361/201833323}, \href
  {https://ui.adsabs.harvard.edu/abs/2019A&A...621A..39E} {621, A39}

\bibitem[\protect\citeauthoryear{{Fabian}, {Walker}, {Russell}, {Pinto},
  {Sanders}  \& {Reynolds}}{{Fabian} et~al.}{2017}]{2017MNRAS.464L...1F}
{Fabian} A.~C.,  {Walker} S.~A.,  {Russell} H.~R.,  {Pinto} C.,  {Sanders}
  J.~S.,   {Reynolds} C.~S.,  2017, \mn@doi [\mnras] {10.1093/mnrasl/slw170},
  \href {https://ui.adsabs.harvard.edu/abs/2017MNRAS.464L...1F} {464, L1}

\bibitem[\protect\citeauthoryear{{Fryxell} et~al.,}{{Fryxell}
  et~al.}{2000}]{2000ApJS..131..273F}
{Fryxell} B.,  et~al., 2000, \mn@doi [\apjs] {10.1086/317361}, \href
  {https://ui.adsabs.harvard.edu/abs/2000ApJS..131..273F} {131, 273}

\bibitem[\protect\citeauthoryear{{Gaspari}, {Ruszkowski}, {Oh}, {Churazov},
  {Brighenti}, {Ettori}, {Sharma}  \& {Temi}}{{Gaspari}
  et~al.}{2013a}]{2013sncl.confE..88G}
{Gaspari} M.,  {Ruszkowski} M.,  {Oh} S.~P.,  {Churazov} E.,  {Brighenti} F.,
  {Ettori} S.,  {Sharma} P.,   {Temi} P.,  2013a, in {Markevitch} M.,  ed.,
  SnowCLUSTER 2013, Physics of Galaxy Clusters. p.~88

\bibitem[\protect\citeauthoryear{{Gaspari}, {Brighenti}  \&
  {Ruszkowski}}{{Gaspari} et~al.}{2013b}]{2013AN....334..394G}
{Gaspari} M.,  {Brighenti} F.,   {Ruszkowski} M.,  2013b, \mn@doi
  [Astronomische Nachrichten] {10.1002/asna.201211865}, \href
  {https://ui.adsabs.harvard.edu/abs/2013AN....334..394G} {334, 394}

\bibitem[\protect\citeauthoryear{{Gaspari}, {Churazov}, {Nagai}, {Lau}  \&
  {Zhuravleva}}{{Gaspari} et~al.}{2014}]{2014A&A...569A..67G}
{Gaspari} M.,  {Churazov} E.,  {Nagai} D.,  {Lau} E.~T.,   {Zhuravleva} I.,
  2014, \mn@doi [\aap] {10.1051/0004-6361/201424043}, \href
  {https://ui.adsabs.harvard.edu/abs/2014A&A...569A..67G} {569, A67}

\bibitem[\protect\citeauthoryear{{Gaspari} et~al.,}{{Gaspari}
  et~al.}{2018}]{2018ApJ...854..167G}
{Gaspari} M.,  et~al., 2018, \mn@doi [\apj] {10.3847/1538-4357/aaaa1b}, \href
  {https://ui.adsabs.harvard.edu/abs/2018ApJ...854..167G} {854, 167}

\bibitem[\protect\citeauthoryear{{Hitomi Collaboration} et~al.,}{{Hitomi
  Collaboration} et~al.}{2016}]{2016Natur.535..117H}
{Hitomi Collaboration} et~al., 2016, \mn@doi [\nat] {10.1038/nature18627},
  \href {https://ui.adsabs.harvard.edu/abs/2016Natur.535..117H} {535, 117}

\bibitem[\protect\citeauthoryear{{Khatri} \& {Gaspari}}{{Khatri} \&
  {Gaspari}}{2016}]{2016MNRAS.463..655K}
{Khatri} R.,  {Gaspari} M.,  2016, \mn@doi [\mnras] {10.1093/mnras/stw2027},
  \href {https://ui.adsabs.harvard.edu/abs/2016MNRAS.463..655K} {463, 655}

\bibitem[\protect\citeauthoryear{{Kim} \& {Ryu}}{{Kim} \&
  {Ryu}}{2005}]{2005ApJ...630L..45K}
{Kim} J.,  {Ryu} D.,  2005, \mn@doi [\apjl] {10.1086/491600}, \href
  {https://ui.adsabs.harvard.edu/abs/2005ApJ...630L..45K} {630, L45}

\bibitem[\protect\citeauthoryear{{Kolmogorov}}{{Kolmogorov}}{1941}]{1941DoSSR..30..301K}
{Kolmogorov} A.,  1941, Akademiia Nauk SSSR Doklady, \href
  {https://ui.adsabs.harvard.edu/abs/1941DoSSR..30..301K} {30, 301}

\bibitem[\protect\citeauthoryear{{Komatsu} et~al.,}{{Komatsu}
  et~al.}{2011}]{2011ApJS..192...18K}
{Komatsu} E.,  et~al., 2011, \mn@doi [\apjs] {10.1088/0067-0049/192/2/18},
  \href {https://ui.adsabs.harvard.edu/abs/2011ApJS..192...18K} {192, 18}

\bibitem[\protect\citeauthoryear{{Lau}, {Kravtsov}  \& {Nagai}}{{Lau}
  et~al.}{2009}]{2009ApJ...705.1129L}
{Lau} E.~T.,  {Kravtsov} A.~V.,   {Nagai} D.,  2009, \mn@doi [\apj]
  {10.1088/0004-637X/705/2/1129}, \href
  {https://ui.adsabs.harvard.edu/abs/2009ApJ...705.1129L} {705, 1129}

\bibitem[\protect\citeauthoryear{{Markevitch} \& {Vikhlinin}}{{Markevitch} \&
  {Vikhlinin}}{2007}]{2007PhR...443....1M}
{Markevitch} M.,  {Vikhlinin} A.,  2007, \mn@doi [\physrep]
  {10.1016/j.physrep.2007.01.001}, \href
  {https://ui.adsabs.harvard.edu/abs/2007PhR...443....1M} {443, 1}

\bibitem[\protect\citeauthoryear{{Miniati}}{{Miniati}}{2015}]{2015ApJ...800...60M}
{Miniati} F.,  2015, \mn@doi [\apj] {10.1088/0004-637X/800/1/60}, \href
  {https://ui.adsabs.harvard.edu/abs/2015ApJ...800...60M} {800, 60}

\bibitem[\protect\citeauthoryear{{Mohapatra}, {Federrath}  \&
  {Sharma}}{{Mohapatra} et~al.}{2020}]{2020MNRAS.493.5838M}
{Mohapatra} R.,  {Federrath} C.,   {Sharma} P.,  2020, \mn@doi [\mnras]
  {10.1093/mnras/staa711}, \href
  {https://ui.adsabs.harvard.edu/abs/2020MNRAS.493.5838M} {493, 5838}

\bibitem[\protect\citeauthoryear{{Mohapatra}, {Federrath}  \&
  {Sharma}}{{Mohapatra} et~al.}{2021}]{2021MNRAS.500.5072M}
{Mohapatra} R.,  {Federrath} C.,   {Sharma} P.,  2021, \mn@doi [\mnras]
  {10.1093/mnras/staa3564}, \href
  {https://ui.adsabs.harvard.edu/abs/2021MNRAS.500.5072M} {500, 5072}

\bibitem[\protect\citeauthoryear{{Nelson}, {Lau}  \& {Nagai}}{{Nelson}
  et~al.}{2014}]{2014ApJ...792...25N}
{Nelson} K.,  {Lau} E.~T.,   {Nagai} D.,  2014, \mn@doi [\apj]
  {10.1088/0004-637X/792/1/25}, \href
  {https://ui.adsabs.harvard.edu/abs/2014ApJ...792...25N} {792, 25}

\bibitem[\protect\citeauthoryear{{Olivares} et~al.,}{{Olivares}
  et~al.}{2019}]{2019A&A...631A..22O}
{Olivares} V.,  et~al., 2019, \mn@doi [\aap] {10.1051/0004-6361/201935350},
  \href {https://ui.adsabs.harvard.edu/abs/2019A&A...631A..22O} {631, A22}

\bibitem[\protect\citeauthoryear{{Randall} et~al.,}{{Randall}
  et~al.}{2015}]{2015ApJ...805..112R}
{Randall} S.~W.,  et~al., 2015, \mn@doi [\apj] {10.1088/0004-637X/805/2/112},
  \href {https://ui.adsabs.harvard.edu/abs/2015ApJ...805..112R} {805, 112}

\bibitem[\protect\citeauthoryear{{Rasia}, {Meneghetti}  \& {Ettori}}{{Rasia}
  et~al.}{2013}]{2013AstRv...8a..40R}
{Rasia} E.,  {Meneghetti} M.,   {Ettori} S.,  2013, \mn@doi [The Astronomical
  Review] {10.1080/21672857.2013.11519713}, \href
  {https://ui.adsabs.harvard.edu/abs/2013AstRv...8a..40R} {8, 40}

\bibitem[\protect\citeauthoryear{{Rebusco}, {Churazov}, {B{\"o}hringer}  \&
  {Forman}}{{Rebusco} et~al.}{2006}]{2006MNRAS.372.1840R}
{Rebusco} P.,  {Churazov} E.,  {B{\"o}hringer} H.,   {Forman} W.,  2006,
  \mn@doi [\mnras] {10.1111/j.1365-2966.2006.10977.x}, \href
  {https://ui.adsabs.harvard.edu/abs/2006MNRAS.372.1840R} {372, 1840}

\bibitem[\protect\citeauthoryear{{Roediger}, {Kraft}, {Nulsen}, {Churazov},
  {Forman}, {Br{\"u}ggen}  \& {Kokotanekova}}{{Roediger}
  et~al.}{2013}]{2013MNRAS.436.1721R}
{Roediger} E.,  {Kraft} R.~P.,  {Nulsen} P.,  {Churazov} E.,  {Forman} W.,
  {Br{\"u}ggen} M.,   {Kokotanekova} R.,  2013, \mn@doi [\mnras]
  {10.1093/mnras/stt1691}, \href
  {https://ui.adsabs.harvard.edu/abs/2013MNRAS.436.1721R} {436, 1721}

\bibitem[\protect\citeauthoryear{{Roncarelli}, {Ettori}, {Borgani}, {Dolag},
  {Fabjan}  \& {Moscardini}}{{Roncarelli} et~al.}{2013}]{2013MNRAS.432.3030R}
{Roncarelli} M.,  {Ettori} S.,  {Borgani} S.,  {Dolag} K.,  {Fabjan} D.,
  {Moscardini} L.,  2013, \mn@doi [\mnras] {10.1093/mnras/stt654}, \href
  {https://ui.adsabs.harvard.edu/abs/2013MNRAS.432.3030R} {432, 3030}

\bibitem[\protect\citeauthoryear{{Russell} et~al.,}{{Russell}
  et~al.}{2019}]{2019MNRAS.490.3025R}
{Russell} H.~R.,  et~al., 2019, \mn@doi [\mnras] {10.1093/mnras/stz2719}, \href
  {https://ui.adsabs.harvard.edu/abs/2019MNRAS.490.3025R} {490, 3025}

\bibitem[\protect\citeauthoryear{{Sanders} \& {Fabian}}{{Sanders} \&
  {Fabian}}{2012}]{2012MNRAS.421..726S}
{Sanders} J.~S.,  {Fabian} A.~C.,  2012, \mn@doi [\mnras]
  {10.1111/j.1365-2966.2011.20348.x}, \href
  {https://ui.adsabs.harvard.edu/abs/2012MNRAS.421..726S} {421, 726}

\bibitem[\protect\citeauthoryear{{Sanders}, {Fabian}, {Smith}  \&
  {Peterson}}{{Sanders} et~al.}{2010}]{2010MNRAS.402L..11S}
{Sanders} J.~S.,  {Fabian} A.~C.,  {Smith} R.~K.,   {Peterson} J.~R.,  2010,
  \mn@doi [\mnras] {10.1111/j.1745-3933.2009.00789.x}, \href
  {https://ui.adsabs.harvard.edu/abs/2010MNRAS.402L..11S} {402, L11}

\bibitem[\protect\citeauthoryear{{Schuecker}, {Finoguenov}, {Miniati},
  {B{\"o}hringer}  \& {Briel}}{{Schuecker} et~al.}{2004}]{2004A&A...426..387S}
{Schuecker} P.,  {Finoguenov} A.,  {Miniati} F.,  {B{\"o}hringer} H.,   {Briel}
  U.~G.,  2004, \mn@doi [\aap] {10.1051/0004-6361:20041039}, \href
  {https://ui.adsabs.harvard.edu/abs/2004A&A...426..387S} {426, 387}

\bibitem[\protect\citeauthoryear{{Shi} \& {Zhang}}{{Shi} \&
  {Zhang}}{2019}]{2019MNRAS.487.1072S}
{Shi} X.,  {Zhang} C.,  2019, \mn@doi [\mnras] {10.1093/mnras/stz1392}, \href
  {https://ui.adsabs.harvard.edu/abs/2019MNRAS.487.1072S} {487, 1072}

\bibitem[\protect\citeauthoryear{{Valdarnini}}{{Valdarnini}}{2019}]{2019ApJ...874...42V}
{Valdarnini} R.,  2019, \mn@doi [\apj] {10.3847/1538-4357/ab0964}, \href
  {https://ui.adsabs.harvard.edu/abs/2019ApJ...874...42V} {874, 42}

\bibitem[\protect\citeauthoryear{{Vall{\'e}s-P{\'e}rez}, {Planelles}  \&
  {Quilis}}{{Vall{\'e}s-P{\'e}rez} et~al.}{2021}]{2021MNRAS.tmp..871V}
{Vall{\'e}s-P{\'e}rez} D.,  {Planelles} S.,   {Quilis} V.,  2021, \mn@doi
  [\mnras] {10.1093/mnras/stab880}, \href
  {https://ui.adsabs.harvard.edu/abs/2021MNRAS.tmp..871V} {}

\bibitem[\protect\citeauthoryear{{Vazza}, {Brunetti}, {Kritsuk}, {Wagner},
  {Gheller}  \& {Norman}}{{Vazza} et~al.}{2009}]{2009A&A...504...33V}
{Vazza} F.,  {Brunetti} G.,  {Kritsuk} A.,  {Wagner} R.,  {Gheller} C.,
  {Norman} M.,  2009, \mn@doi [\aap] {10.1051/0004-6361/200912535}, \href
  {https://ui.adsabs.harvard.edu/abs/2009A&A...504...33V} {504, 33}

\bibitem[\protect\citeauthoryear{{Vazza}, {Roncarelli}, {Ettori}  \&
  {Dolag}}{{Vazza} et~al.}{2011a}]{2011MNRAS.413.2305V}
{Vazza} F.,  {Roncarelli} M.,  {Ettori} S.,   {Dolag} K.,  2011a, \mn@doi
  [\mnras] {10.1111/j.1365-2966.2010.18120.x}, \href
  {https://ui.adsabs.harvard.edu/abs/2011MNRAS.413.2305V} {413, 2305}

\bibitem[\protect\citeauthoryear{{Vazza}, {Brunetti}, {Gheller}, {Brunino}  \&
  {Br{\"u}ggen}}{{Vazza} et~al.}{2011b}]{2011A&A...529A..17V}
{Vazza} F.,  {Brunetti} G.,  {Gheller} C.,  {Brunino} R.,   {Br{\"u}ggen} M.,
  2011b, \mn@doi [\aap] {10.1051/0004-6361/201016015}, \href
  {https://ui.adsabs.harvard.edu/abs/2011A&A...529A..17V} {529, A17}

\bibitem[\protect\citeauthoryear{{Vazza}, {Roediger}  \& {Br{\"u}ggen}}{{Vazza}
  et~al.}{2012}]{2012A&A...544A.103V}
{Vazza} F.,  {Roediger} E.,   {Br{\"u}ggen} M.,  2012, \mn@doi [\aap]
  {10.1051/0004-6361/201118688}, \href
  {https://ui.adsabs.harvard.edu/abs/2012A&A...544A.103V} {544, A103}

\bibitem[\protect\citeauthoryear{{Vazza}, {Jones}, {Br{\"u}ggen}, {Brunetti},
  {Gheller}, {Porter}  \& {Ryu}}{{Vazza} et~al.}{2017}]{2017MNRAS.464..210V}
{Vazza} F.,  {Jones} T.~W.,  {Br{\"u}ggen} M.,  {Brunetti} G.,  {Gheller} C.,
  {Porter} D.,   {Ryu} D.,  2017, \mn@doi [\mnras] {10.1093/mnras/stw2351},
  \href {https://ui.adsabs.harvard.edu/abs/2017MNRAS.464..210V} {464, 210}

\bibitem[\protect\citeauthoryear{{Vazza}, {Angelinelli}, {Jones}, {Eckert},
  {Br{\"u}ggen}, {Brunetti}  \& {Gheller}}{{Vazza}
  et~al.}{2018}]{2018MNRAS.481L.120V}
{Vazza} F.,  {Angelinelli} M.,  {Jones} T.~W.,  {Eckert} D.,  {Br{\"u}ggen} M.,
   {Brunetti} G.,   {Gheller} C.,  2018, \mn@doi [\mnras]
  {10.1093/mnrasl/sly172}, \href
  {https://ui.adsabs.harvard.edu/abs/2018MNRAS.481L.120V} {481, L120}

\bibitem[\protect\citeauthoryear{{Wang} \& {Markevitch}}{{Wang} \&
  {Markevitch}}{2018}]{2018ApJ...868...45W}
{Wang} Q. H.~S.,  {Markevitch} M.,  2018, \mn@doi [\apj]
  {10.3847/1538-4357/aae921}, \href
  {https://ui.adsabs.harvard.edu/abs/2018ApJ...868...45W} {868, 45}

\bibitem[\protect\citeauthoryear{{Wittor}, {Jones}, {Vazza}  \&
  {Br{\"u}ggen}}{{Wittor} et~al.}{2017}]{2017MNRAS.471.3212W}
{Wittor} D.,  {Jones} T.,  {Vazza} F.,   {Br{\"u}ggen} M.,  2017, \mn@doi
  [\mnras] {10.1093/mnras/stx1769}, \href
  {https://ui.adsabs.harvard.edu/abs/2017MNRAS.471.3212W} {471, 3212}

\bibitem[\protect\citeauthoryear{{Zhuravleva} et~al.,}{{Zhuravleva}
  et~al.}{2014a}]{2014Natur.515...85Z}
{Zhuravleva} I.,  et~al., 2014a, \mn@doi [\nat] {10.1038/nature13830}, \href
  {https://ui.adsabs.harvard.edu/abs/2014Natur.515...85Z} {515, 85}

\bibitem[\protect\citeauthoryear{{Zhuravleva} et~al.,}{{Zhuravleva}
  et~al.}{2014b}]{2014ApJ...788L..13Z}
{Zhuravleva} I.,  et~al., 2014b, \mn@doi [\apjl] {10.1088/2041-8205/788/1/L13},
  \href {https://ui.adsabs.harvard.edu/abs/2014ApJ...788L..13Z} {788, L13}

\bibitem[\protect\citeauthoryear{{Zhuravleva}, {Allen}, {Mantz}  \&
  {Werner}}{{Zhuravleva} et~al.}{2018}]{2018ApJ...865...53Z}
{Zhuravleva} I.,  {Allen} S.~W.,  {Mantz} A.,   {Werner} N.,  2018, \mn@doi
  [\apj] {10.3847/1538-4357/aadae3}, \href
  {https://ui.adsabs.harvard.edu/abs/2018ApJ...865...53Z} {865, 53}

\makeatother
\end{thebibliography}
\newpage

\setcounter{figure}{0}
\renewcommand{\figurename}{Fig.}
\renewcommand{\thefigure}{A\arabic{figure}}
\clearpage
\section{Appendix 1: On the $Ri - \sigma^2_s$ relation at different filtering scales}

In Sec.~\ref{sec:Ri} we showed that the Richardson number is very dependent on the filtering scale. In order to validate the robustness of our results, we performed tests with a filtering scale of 100 and 800 kpc. Due to the resolution of our simulation, an analysis with a much smaller scale cannot be accomplished.

\begin{figure}[h!]
        \centering
        \includegraphics[width= 0.5\textwidth]{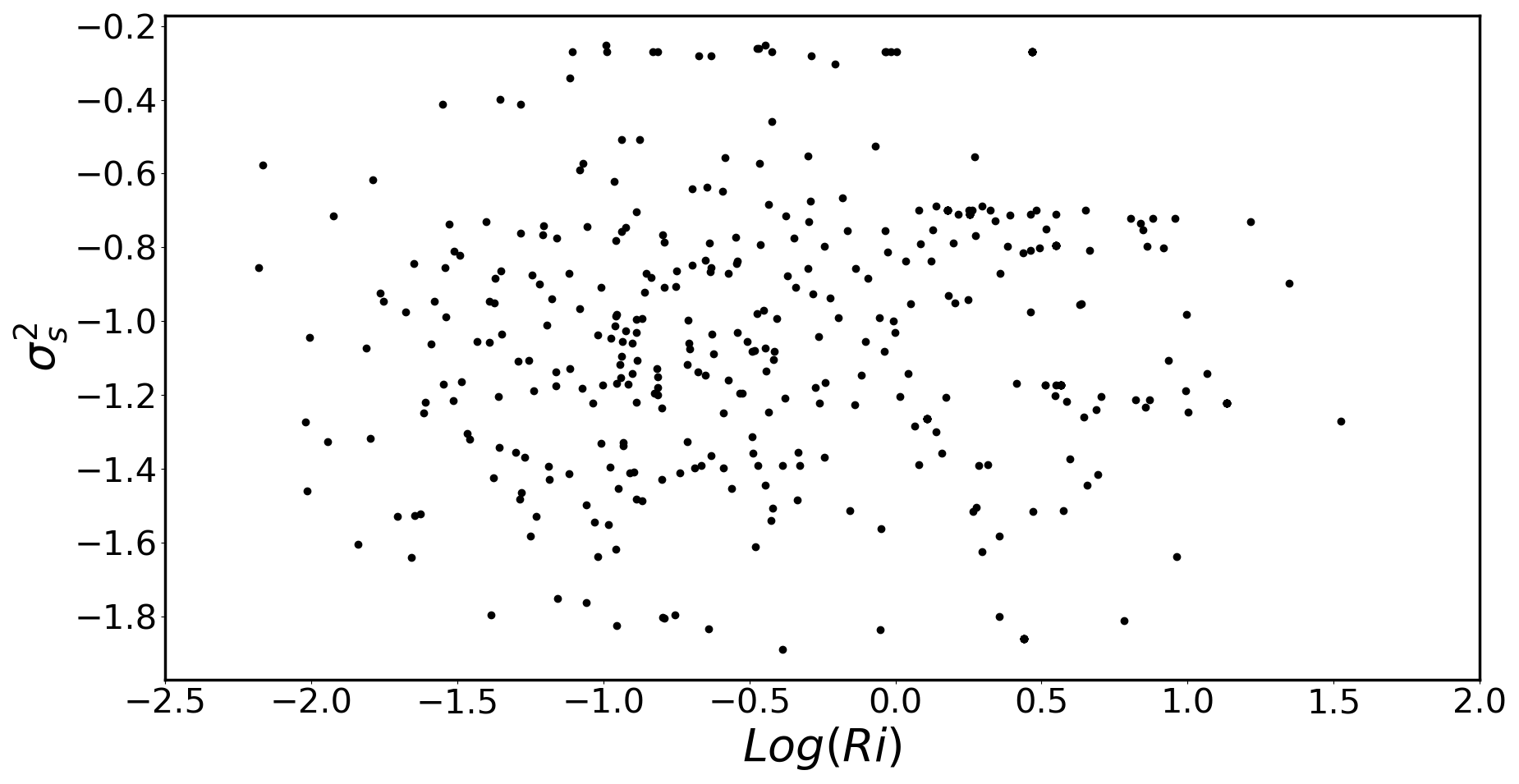}\quad\includegraphics[width= 0.5\textwidth, height=0.3\textwidth]{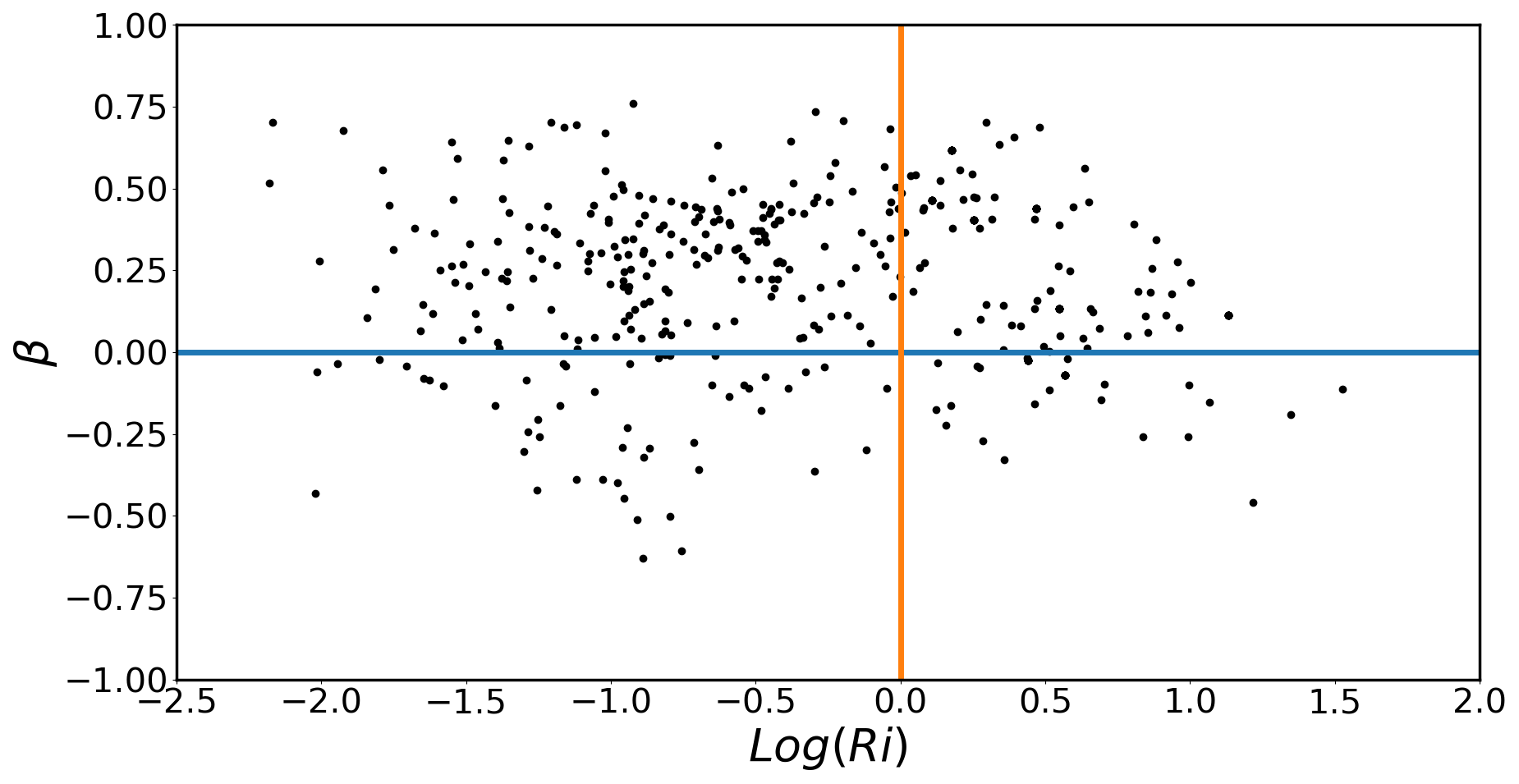}
        \caption {$Ri - \sigma^2_s$ and $Ri - \beta$ plot for a filtering scale $L = 100 $ kpc. The orange line is used to highlight the turbulence (left) and buoyancy (right) dominated regime, while the blue line distinguishes a radially ($\beta > 0$) from a tangentially ($\beta < 0$) bias.
}
        \label{fig:Ri_beta_100}
\end{figure}

\begin{figure}
        \centering
        \includegraphics[width= 0.5\textwidth]{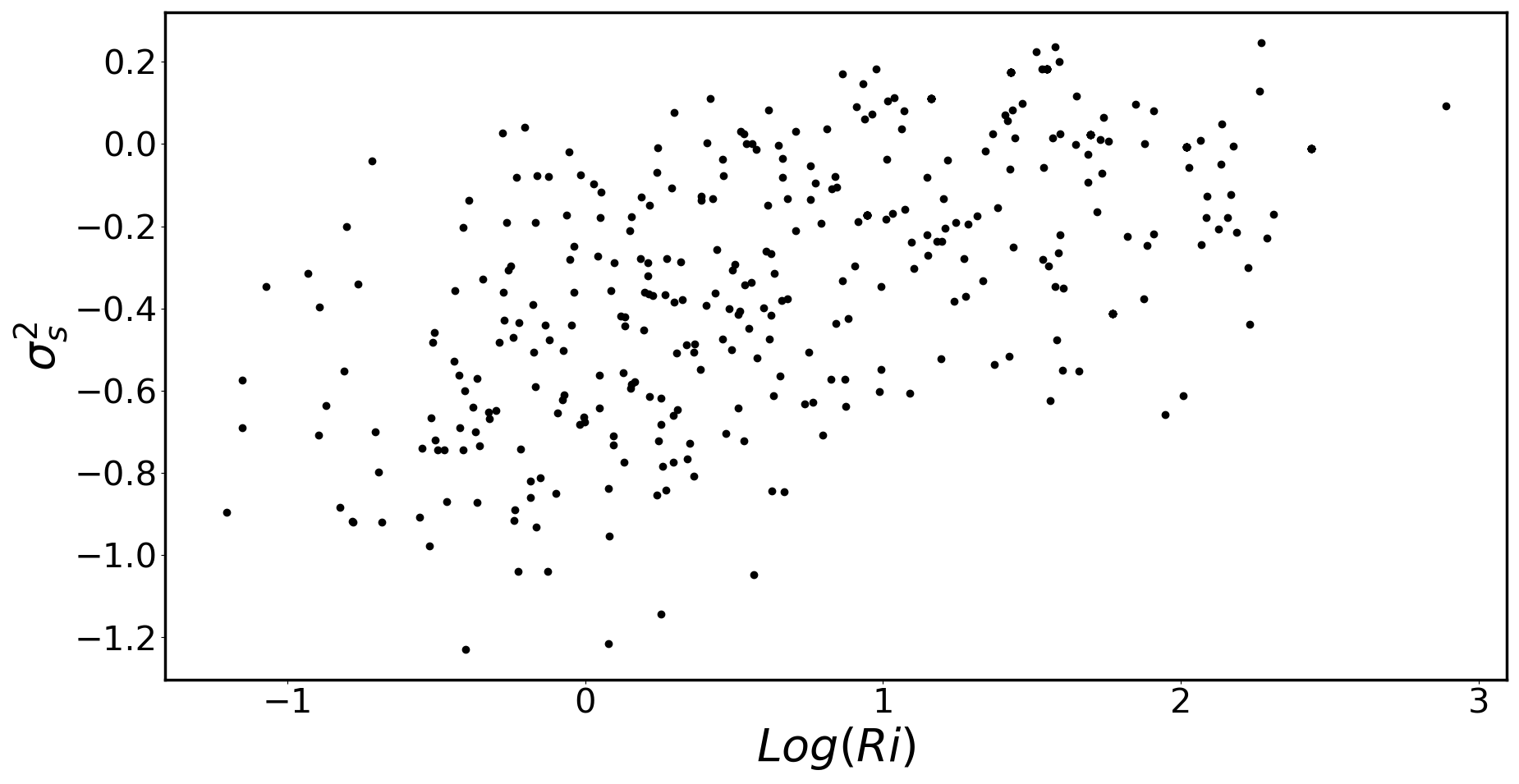}\quad\includegraphics[width= 0.5\textwidth]{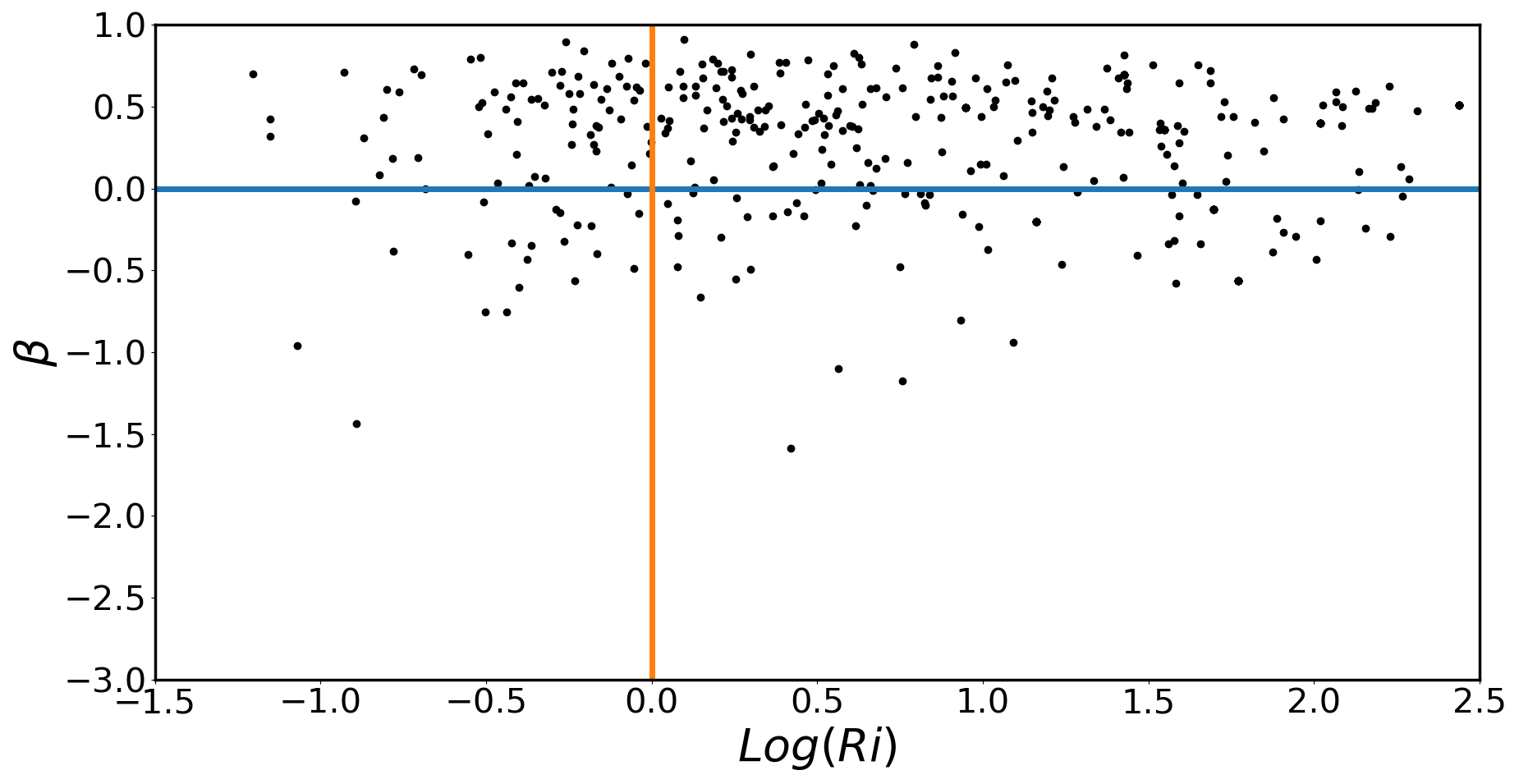}
        \caption {$Ri - \sigma^2_s$ and $Ri - \beta$ plot for a filtering scale $L = 800 $ kpc. The orange line is used to highlight the turbulence (left) and buoyancy (right) dominated regime, while the blue line distinguishes a radially ($\beta > 0$) from a tangentially ($\beta < 0$) bias.}
        \label{fig:Ri_beta_800}
\end{figure}

Fig.~\ref{fig:Ri_beta_100} and \ref{fig:Ri_beta_800} show the $Ri- \sigma^2_s$ and $Ri- \beta$ relation for $L = 100$ kpc  and $L = 800$ kpc, respectively.  In essence, the results are  almost identical to those of Sec.~\ref{sec:Ri_sigma}, as we did not detect any strong relation between the variables. It looks there is a $Ri - \sigma^2_s$ relation for $ L = 800$ kpc. However, there is a lot a scatter and the fit would definitely be different from that of \citeauthor{2020MNRAS.493.5838M} since the density fluctuations are much higher. Furthermore, a filtering algorithm with such a large scale, it might not be a properly choice, since we are essentially removing the bulk motions on scale of typical size of the cluster virial radius, leaving out the laminar flows on smaller scales. Besides, the bottom panel of Fig. \ref{fig:Ri_beta_800} shows that a radial bias is still present, confirming that the Richardson number is not the ideal proxy to quantify the effect of stratification on turbulence in galaxy clusters.

\end{document}